\documentstyle[epsf]{mn}
\title[H$\alpha$ Survey of Abell Clusters]
{An H$\alpha$ Survey of 8 Abell Clusters : 
the dependence of tidally-induced star formation on cluster density}
\author[C.~Moss and M.~Whittle]
{C.~Moss,$^1$\thanks{Email: cmm@astro.livjm.ac.uk}\thanks{Present address:
Astrophysics Research Institute, Liverpool John Moores University,
Twelve Quays House, Egerton Wharf, Birkenhead CH41 1LD}
and M.~Whittle$^2$\\
$^{1}$Vatican Observatory Research Group, Steward Observatory, University of Arizona,
Tucson, AZ 85721, USA \\
$^{2}$Department of Astronomy, University of Virginia, Charlottesville, VA
22903, USA
}
\date{22 February 2000}
\pubyear{2000}
\begin{document}
\maketitle

\begin{abstract}

We have undertaken a survey of H$\alpha$ emission in a substantially
complete sample of CGCG galaxies of types Sa and later within 1.5 Abell radii
of the centres of 8 low-redshift Abell clusters (Abell 262, 347, 400, 426,
569, 779, 1367 and 1656).  Some 320 galaxies were surveyed, of which 116
were detected in emission (39\% of spirals, 75\% of peculiars).
Here we present previously unpublished data
for 243 galaxies in 7 clusters.

Detected emission was classified as `compact' or `diffuse'. From an 
analysis of the full survey sample, we reconfirm our previous identification
of compact and diffuse emission with circumnuclear starburst and disk
emission respectively. The circumnuclear emission is associated either with
the presence of a bar, or with a disturbed galaxy morphology indicative of
on-going tidal interactions (whether galaxy--galaxy, galaxy--group, or
galaxy--cluster).

The frequency of such tidally-induced (circumnuclear) starburst
emission in spirals increases from regions of lower to higher local
galaxy surface density, and from clusters with lower to higher central
galaxy space density. The percentages of spirals classed as disturbed,
and of galaxies classified as peculiar show a similar trend.  These
results suggest that tidal interactions for spirals are more frequent
in regions of higher local density and for clusters with higher
central galaxy density.  The prevalence of such tidal interactions in
clusters is expected from recent theoretical modelling of clusters
with a non-static potential undergoing collapse and infall.
Furthermore, in accord with this picture, we suggest that peculiar
galaxies are predominantly on-going mergers.

We conclude that tidal interactions are likely to be the main
mechanism for the transformation of spirals to S0s in clusters.  This
mechanism operates more efficiently in higher density environments as
is required by the morphological type--local surface density
(T--$\Sigma$) relation for galaxies in clusters.  For regions of
comparable local density, the frequency of tidally-induced starburst
emission is greater in clusters with higher central galaxy density.
This implies that, for a given local density, morphological
transformation of disk galaxies proceeds more rapidly in clusters of
higher central galaxy density. This effect is considered to be due to
subcluster merging and could account for the previously considered
anomalous absence of a significant T--$\Sigma$ relation for irregular
clusters at intermediate redshift.

\end{abstract}


\section {Introduction}
\label{intro}

The systematic differences in morphology between field and cluster
galaxy populations have long been known (e.g. Hubble \& Humason 1931;
Oemler 1974). More recently, data from HST has shown the remarkable
changes in cluster galaxy populations between intermediate redshifts
($z\sim0.5$) and the present.  Intermediate redshift clusters contain
a large population of blue, star forming galaxies, which have been
shown to be predominantly normal spiral and irregular galaxies, a
fraction of which are interacting or obviously disturbed
(e.g. Butcher \& Oemler 1978; Dressler et al. 1994; Oemler, Dressler
\& Butcher 1997; Smail et al. 1997). They constitute up to
50\% of the cluster population, but by the present epoch have been
depleted by a factor of two in rich clusters and have been replaced by a
corresponding increase in the S0 population (Oemler 1974; Dressler
1980; Dressler et al. 1997).  What processes are responsible for this
rapid depletion of the spiral population and corresponding increase in
S0s in rich clusters since $z=0.5$?  There have been many suggested
mechanisms, either to remove gas and/or induce star formation, some of
which depend on galaxy--galaxy collisions (e.g. Spitzer \& Baade 1951;
Miller 1988; Valluri \& Jog 1990) or on the effect of the intracluster medium
(e.g. Gunn \& Gott 1972; Cowie \& Songaila 1977), or on tidal shocks
whether from galaxy--galaxy or cluster--galaxy interactions (e.g.
Noguchi \& Ishibashi 1986; Lavery \& Henry 1988; Sanders et al. 1988; Henriksen
\& Byrd 1996; Moore et al. 1996).

Nearby rich clusters have a residual population of spiral galaxies.
If one or several of the proposed mechanisms have been operating to
transform spirals into S0s over the relatively short look-back time to
$z=0.5$, it is clear that we might expect the same processes to be
continuing to operate in the present on the residual population of
spirals in clusters.  These processes can be more easily studied in
nearby clusters than at higher redshifts.  Furthermore all of the
proposed mechanisms involve potentially dramatic changes in the star
formation rates in spirals.  Thus a comparison of star formation rates
between spirals in nearby clusters and those in the field may provide
the observational evidence to help decide the physical mechanism which
has been responsible for the dramatic recent change in the cluster
disk galaxy population.

In practice, it has proved difficult to establish agreement amongst
different authors regarding changes of star formation rate between
field and cluster spirals.  However, much recent work supports either
similar or enhanced star formation in cluster spirals compared to
field spirals (e.g. Donas et al. 1990; Moss \& Whittle 1993 [Paper
II]; Gavazzi \& Contursi 1994; Biviano et al. 1997; Moss, Whittle \&
Pesce 1998 [Paper III]; Gavazzi et al. 1998). Biviano et al. have
suggested that earlier studies which claimed reduced star formation in
cluster spirals may have been affected by an unrecognised bias whereby
faint field galaxies are more likely to be detected in emission than
their cluster counterparts.  Two recent studies (Balogh et al. 1998;
Hashimoto et al. 1998) have found a suppression of star formation in
cluster galaxies relative to galaxies of similar morphological type in
the field.  However the morphological classifications in these studies
are based on bulge-to-disk ratio, and it is not clear to what extent
the results are affected by the variation of S0/S ratio from the field
to the cluster (see section \ref{cfcspirals} below for further
discussion).  Furthermore it is also increasingly evident that star
formation in the spiral disks and in the circumnuclear region may have
very different dependencies on environment (cf. Moss et al. 1998;
Hashimoto et al. 1998).

We have made an extensive survey of H$\alpha$ emission as an indicator
of the star formation rate in spirals in nearby clusters (Paper III
and references therein). One motivation is to understand
how the cluster environment affects the evolution of spiral galaxies,
including the dramatic depletion of cluster spirals over the past few
giga-years. Our survey technique can distinguish well between disk
emission and circumnuclear starburst emission, and accordingly
investigate how these vary with environment.  In previous work we have
discussed in detail a comparison between emission in field spirals
and a single cluster, Abell 1367 (Paper III).  Here we utilise
data for all 8 clusters in our sample, and do a comparable
analysis for a full range of cluster types, discussing how emission
varies across a range of environments of differing galaxy densities.
We also attempt to differentiate the dependence of emission on local
galaxy density from that on cluster type to give further insight into
evolutionary mechanisms operating on cluster spirals.

The paper is set out as follows.  In section \ref{obs} we describe the
survey sample and summarise observational and emission detection
methods. A previously unpublished list of emission-line galaxies
(ELGs) detected for 6 of the 8 surveyed clusters is given in section
\ref{gprops}. In this section we also consider the relation of
emission to galaxy properties, and show that compact and diffuse
emission detected on the prism plates can be well understood as
circumnuclear starburst and normal disk emission respectively.  In
section \ref{cenviron}, using a variety of cluster/field parameters,
we show that there is a systematic enhancement of tidally-induced
starburst emission with increasingly rich clusters.  For the richest
clusters this enhancement is greater than would be expected simply on
the basis of increasing galaxy density alone.  These observational
results are discussed in section \ref{discuss}, where we show that
they provide convincing evidence that spirals have been transformed to
S0s in clusters predominantly by tidal forces, a picture fully in accord
with the most recent numerical simulations of clusters (e.g.  Gnedin
1999).  We further discuss how the observational results can explain
the apparently anomalous result for type--galaxy surface density relation found
by Dressler et al. (1997) for low richness clusters at intermediate
redshifts.  A summary of our results is given in section
\ref{conclude}.

\section {Observations and Measurements}
\label{obs}

\subsection {Cluster and galaxy samples}
\label{sample}

\begin{figure*}

\epsfxsize=12cm
\hspace*{0.01cm} \epsffile{fig1a.epsi}

\caption{\label{fields} CGCG galaxies in cluster fields. Galaxy symbols are 
the same as
Zwicky et al. (1960--1968): cross superimposed on a filled square,
\mbox{$m_{\rm p}$ $\le$ 11.0}; filled square, 
\mbox{$m_{\rm p}$ = 11.1 -- 12.0};
open square, \mbox{$m_{\rm p}$ = 12.1 -- 13.0};
filled circle, \mbox{$m_{\rm p}$ = 13.1 -- 14.0};
open circle, \mbox{$m_{\rm p}$ = 14.1 -- 15.0}; open triangle, 
\mbox{$m_{\rm p}$ = 15.1 -- 15.7}.
Plate boundaries are shown schematically with dashed lines.
The solid and dotted lines are circles of radius, $r$ = 1.5 ${\rm r}_{\rm A}$
and 3.0 ${\rm r}_{\rm A}$ 
respectively centred on the cluster centre. Note that CGCG galaxies are
only shown for \mbox{$r \le 3.0$ ${\rm r}_{\rm A}$}. }

\end{figure*}

\setcounter{figure}{0}

Table \ref{clusters} gives basic data for the 8 Abell clusters in our
survey (Abell 262, 347, 400, 426, 569, 779, 1367, and 1656).  These
clusters constitute a representative sample, comprising all but two of
the 10 Abell clusters in the northern hemisphere with redshifts less
than $7200$ km s$^{-1}$ (the other two clusters, Abell 189 and 194, are
both relatively poor clusters comparable to Abell 262, 347, 569 and
779).

Our initial sample of galaxies comprised all CGCG galaxies (Zwicky et
al. 1960--1968) within 1.5 Abell radii of the cluster centres (759
galaxies, where resolved double galaxies are counted as two). These
galaxies were morphologically classified (see section \ref{types}) and
a subset defined which excluded galaxies with Hubble types E, E/S0, S0,
S0/a or galaxies of indeterminate type (292 galaxies remaining). A
further 28 spirals falling beyond 1.5 Abell radii were included (27 in
Abell 1367, 1 in Abell 400), yielding a final total of 320 galaxies
selected for the survey for H$\alpha$ emission.  Our restriction to
CGCG galaxies reflects the fact that our detection efficiency decreases
sharply below the CGCG magnitude limit  $m_{\rm p} = 15.7$, and our
exclusion of E, E/S0, S0, S0/a galaxies reflects the fact that in
practice these Hubble types are rarely detected in H$\alpha$ (see Paper
III). In the case of double galaxies, those 11 individual members
fainter than 15.7 were excluded from the statistical sample, as were 15
galaxies which, for various reasons, were only visible on one of our
two plates. Thus, our final statistical sample represents a
substantially complete group of potentially detectable star-forming
galaxies in and around nearby Abell clusters.

Cluster centres are taken from Abell, Corwin \& Olowin (1989). Cluster mean
redshifts, $z_{o}$, and velocity dispersions, $\sigma_{v}$, based on a
total of $n$ redshifts, are taken from Struble \& Rood (1991), where
$z_{o}$ has been corrected to the centroid of the Local Group following RC2
(de Vaucouleurs, de Vaucouleurs \& Corwin 1976).  The Abell radius is
defined (Abell 1958) as $5.13\times10^5/cz_{o}$ arcmin and corresponds to
$\sim 1.5 h^{-1}$ Mpc where $h$ is the Hubble constant in units of 100
km s$^{-1}$ Mpc$^{-1}$.

\subsection {Plate material and H$\alpha$ detection}
\label{pplates}

\begin{figure}

\epsfxsize=5cm
\hspace*{1.5cm} \epsffile{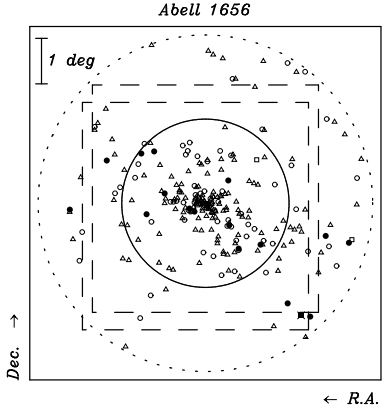}

\caption{continued.}
\end{figure}

Table \ref{plates} gives basic information about the objective prism plates
used for the survey, while Figure \ref{fields} shows the distribution of
CGCG galaxies within the Abell clusters as well as the objective prism plate
boundaries (see Paper III for Abell 1367). 

Our survey technique and methods have been described in detail in
Paper I (Moss, Whittle \& Irwin 1988) and to a lesser extent in Papers
II and III. Here, we briefly review the methods.  All plates were
taken on the 61/94 cm Burrell Schmidt telescope at Kitt Peak in
conditions of good seeing and transparency, and are consequently of
good quality.  The plates cover approximately $5\degr$ at 94 arcsec
${\rm mm}^{-1}$ and use an emulsion/filter combination of either
\mbox{IIIaF/RG630(round)} or
\mbox{IIIaF/RG645(square)} giving $\sim 350$ \AA\/ bandpass centered on
6655 \AA\/ with a peak sensitivity $\sim$ 6717 \AA.  Two prisms were
used, either a high dispersion $10\degr$ prism giving $\sim 400$ \AA\/
mm$^{-1}$, or, when this became unavailable, a lower dispersion
\mbox{$2\degr\! + 4\degr$} prism combination giving $\sim 780$ \AA\/
mm$^{-1}$.  In Paper III we compared the H$\alpha$ detection efficiency
of these two prism combinations and concluded that they were
substantially equivalent.  Each cluster was observed twice, with the
telescope east and west of the pier to reverse the dispersion
direction. Having two such plates not only ensures a more reliable
detection of H$\alpha$, but also, from the difference in location of
the emission, yields relatively accurate measurements of redshift.

Using a low power binocular microscope ($\sim 12\times$), the galaxy
spectra were inspected for signs of H$\alpha$ emission, which appears
as an H$\alpha$ image superposed on the dispersed continuum spectrum.
In Paper III we analysed the H$\alpha$ sensitivity limit and found that
the objective prism technique is 90\% complete down to an
equivalent width limit of $20$ \AA\/ for the H$\alpha$+[NII] blend, and
$\sim 29\%$ efficient below this limit.

Table \ref{gsample} gives the surveyed galaxies and H$\alpha$ detections
for seven clusters, while Paper III give these for the eighth, Abell 1367.

\subsection  {Parameters and Ranking}
\label{ranks}

Our statistical analysis requires a range of parameters to characterise
galaxy morphology, H$\alpha$ emission, local environment, and more global
environment.  We list these parameters in Table \ref{tparam} together
with their quantification as ranked and/or binned data suitable for
the non-parametric statistical tests used below (sections \ref{mags}
and following), and the sample number, $n$, for each rank or bin. A
more detailed description of individual parameters is as follows.

\subsubsection  {H$\alpha$ Emission}
\label{emission}

For each detected galaxy, the H$\alpha$ emission was graded for
visibility on a 5 point scale (S -- strong; MS -- medium-strong; M --
medium; MW -- medium-weak; and W -- weak).  Similarly, the appearance
of the H$\alpha$ image was classified on a 5 point scale (VC -- very
concentrated; C -- concentrated; N -- normal; D -- diffuse; and VD --
very diffuse).  Concentrated emission is much brighter than the
underlying continuum and is sharply delineated from it; diffuse
emission is only slightly brighter than the continuum and has an
indistinct appearance, and in general spans a larger region than the
concentrated emission (for further discussion, see Paper II).
Galaxies with double, multiple, or offset emission were categorised as
compact emission, principally because of its high surface brightness.
The combined mean emission classifications from each plate pair are
listed in columns 12 and 13 of Table \ref{gsample}.

We choose a binary rank for H$\alpha$ detection, with no rank assigned
for galaxies not satisfactorily surveyed for emission (`?' in Table
\ref{gsample}). We also choose binary ranks for the H$\alpha$
appearance, yielding two parameters: {\it compact} emission
(concentration classes VC, C, or N); and {\it diffuse} emission
(concentration classes D or VD).  
Astrophysically, we associate
compact emission with a circumnuclear starburst, and diffuse emission
with more normal ongoing disk-wide star formation. (For further
discussion, see section \ref{gpdiscuss} below.)

\subsubsection  {Hubble Types}
\label{types}

Because star formation rates depend quite sensitively on Hubble type,
it is important to estimate these types accurately, so that dependence
on environment can be clearly distinguished from dependence on Hubble
type.  Unfortunately, at redshifts of $\sim 6000$ km s$^{-1}$ the
cluster galaxies are quite small and difficult to type accurately
without good plate material.  In the case of Abell 1367 we indeed have
excellent plate material, and so used it to perform the typing of most
of the CGCG galaxies in that cluster (see Paper III). For the other
clusters this is not the case, and we have adopted a somewhat more
conservative approach to galaxy typing. For all UGC (Nilson 1973)
galaxies, we adopt the UGC type. For non-UGC galaxies one of us (MW)
classified the galaxies on the revised de Vaucouleurs (1959, 1974)
system using glass copies of the PSS and direct IIIaJ plates taken on
the Burrell Schmidt.  The reliability of these types was assessed in
two ways.  Firstly, UGC galaxies were also typed and comparison with
UGC types showed a standard deviation in the T class of $\sim 1.0$.
Secondly, all galaxies were independently typed twice and showed
similar level of agreement.

\begin{table*}
\begin{minipage}{130mm}
\caption{Clusters included in the H$\alpha$ survey}
\label{clusters}
\begin{tabular}{lrrrrrrcccr}
Cluster & \multicolumn{6}{c}{Cluster centre} & \multicolumn{1}{c}{Abell radius} & 
\multicolumn{1}{c}{ $z_{o}$ } &
\multicolumn{1}{c}{ $\sigma_{v}$ } & \multicolumn{1}{c}{$n$} \\ \cline{2-7}
 & \multicolumn{4}{c}{R.A. (1950) Dec.} & \multicolumn{1}{c}{$l$} &
\multicolumn{1}{c}{$b$} & \multicolumn{1}{c}{(arcmin)} &&
\multicolumn{1}{c}{(km ${\rm s}^{-1}$)} & \\
 &&&&&&&& \\
 Abell 262  & 
 1$^{\rm h}$ \hspace{-0.85em} &
 49\fm9 \hspace{-0.75em} &
 35$^{\circ}$ \hspace{-0.85em} &
 54$^{\prime}$ \hspace{-0.65em} &
 136\fdg59 \hspace{-0.50em} &
 -25\fdg09 \hspace{-0.50em} &      105 \hspace{0.35em} &     0.0163&    
494  &    47 \\ 
 Abell 347  &    2& 22.7&    41& 39&   141.17&-17.63&       91 &     0.0189&    
582  &    21 \\                                         
 Abell 400  &    2& 55.0&     5& 50&   170.25&-44.93&       72 &     0.0238&    
610  &    71 \\  
 Abell 426  &    3& 15.3&    41& 20&   150.39&-13.38&       96 &     0.0179&   
1277 \hspace{0.20em} &   114 \\                                         
 Abell 569  &    7&  5.4&    48& 42&   168.58& 22.81&       88 &     0.0196&    
444  &    12 \\                                         
 Abell 779  &    9& 16.8&    33& 59&   191.07& 44.41&       75 &     0.0230&    
472  &    24 \\
 Abell 1367 &   11& 41.9&    20&  7&   234.81& 73.03&       80 &     0.0214&    
822  &    93 \\                                         
 Abell 1656 &   12& 57.4&    28& 15&    58.09& 87.96&       74 &     0.0232&    
880  &   226 \\ 
\end{tabular}
\end{minipage}
\end{table*}

\medskip

For 92 galaxies, it was not possible to determine a reliable type due
their small and/or saturated images on the PSS.  In Abell 1367 and
1656, some 41 such galaxies with $m_{\rm p} \le 15.7$ were inspected
for emission on the prism plate pairs for these clusters.  A total of 6
galaxies (15\% of the sample) were found to have emission. This is a
larger percentage emission detection than for early type galaxies
($\sim 5\%$), but much smaller than for galaxies of types Sa and later
($\sim 40\%$). We conclude that the galaxies with indeterminate type
are likely to be predominantly early type galaxies, and accordingly it
was decided also to omit them from the study, as mentioned in section
\ref{sample}.  These galaxies should not be confused with `peculiar'
galaxies, nor with spirals with postfix `pec', both of which have been
retained in the sample.

The Hubble types were reduced to 8 category binned data as shown in
Table \ref{tparam}; and the spiral stage has been given a simple rank
of $1 - 6$.  The rank assigned is independent of bar designation or
whether `pec' is appended to the class, but no rank is assigned if a
spiral has no stage assigned or the galaxy is classed simply as
Peculiar.

    \begin{table*}
    \centering
\caption{\label{plates} Plate material}
 
\begin{tabular}{rllrrrrcrrc} 
\multicolumn{1}{c}{Plate no.} &
\multicolumn{1}{c}{U.T. date} &
\multicolumn{1}{c}{Cluster} &
\multicolumn{4}{c}{Plate centre} &
\multicolumn{1}{c}{Prism} &
\multicolumn{1}{c}{Filter} &
\multicolumn{1}{c}{Exp.} &
\multicolumn{1}{c}{Tel.} \\ \cline{4-7}
 &&& \multicolumn{4}{c}{R.A. (1950) Dec.} &&& \multicolumn{1}{c}{(min.)} &
 \multicolumn{1}{c}{(E/W)}  \\
 &&&&&&&&&& \\
  15204 & 1984 Nov 4 & Abell 262  &
  1$^{\rm h}$ \hspace{-0.85em} &
  50\fm0 \hspace{-0.75em}  &
  35$^{\circ}$ \hspace{-0.85em}  &
  44$^{\prime}$ \hspace{-0.65em} & 2+4 & RG 645 &  60 & E \\
  15205 & 1984 Nov 4 & Abell 262  &  1 & 50.2 & 36 & 13 & 2+4 & RG 645 &  60 & W \\
  13046 & 1981 Dec 16& Abell 347  &  2 & 21.9 & 41 & 20 & 10  & RG 630 & 120 & W \\
  14559 & 1983 Oct 29& Abell 347  &  2 & 24.7 & 41 & 36 & 10  & RG 645 & 120 & E \\
  15198 & 1984 Nov 2 & Abell 400  &  2 & 55.5 &  6 & 23 & 2+4 & RG 645 & 120 & W \\
  15201 & 1984 Nov 3 & Abell 400  &  2 & 56.4 &  5 & 32 & 2+4 & RG 645 &  90 & E \\
  15191 & 1984 Oct 31& Abell 426  &  3 & 16.8 & 41 & 30 & 2+4 & RG 645 & 120 & W \\
  15195 & 1984 Nov 1 & Abell 426  &  3 & 15.7 & 41 & 26 & 2+4 & RG 645 & 120 & E \\
  15196 & 1984 Nov 1 & Abell 569  &  7 &  4.9 & 48 & 28 & 2+4 & RG 645 & 120 & E \\
  15230 & 1984 Dec 31& Abell 569  &  7 &  5.9 & 48 & 57 & 2+4 & RG 645 &  97 & W \\
  14078 & 1983 Apr 4 & Abell 779  &  9 & 17.3 & 33 & 56 & 10  & RG 630 &  70 & E \\
  14193 & 1983 May 1 & Abell 779  &  9 & 16.2 & 33 & 47 & 10  & RG 630 & 120 & W \\
  14077 & 1983 Apr 3 & Abell 1367 & 11 & 37.9 & 19 & 59 & 10  & RG 630 &  75 & E \\
  14200 & 1983 May 3 & Abell 1367 & 11 & 41.9 & 20 & 00 & 10  & RG 630 & 120 & W \\
  15270 & 1985 Apr 11& Abell 1656 & 12 & 58.4 & 27 & 58 & 2+4 & RG 645 & 120 & E \\
  15271 & 1985 Apr 11& Abell 1656 & 12 & 57.4 & 28 & 21 & 2+4 & RG 645 & 120 & W \\
\end{tabular}
    \end{table*}

\subsubsection{Barred structure}
\label{barred}

Since galaxy bars may be related to star formation and environment, we
attempted to assign bar designations for all galaxies following the de
Vaucouleurs system.  Unfortunately, many UGC types do not distinguish
between a non-barred spiral (e.g. SAa) and a spiral with no bar
designation (e.g. Sa).  Accordingly,  we inspected all galaxies for
signs of a bar and, where possible, assigned a bar type (e.g. SA, SAB,
or SB).  This bar classification was included in the type descriptions
of the galaxies in the sample given in column seven of \mbox{Table
\ref{gsample}}.

The bar classifications form a rank sequence ($1-4$, see Table
\ref{tparam}) from unbarred (A, A:), through intermediate (AB, AB:), to
uncertain bar (B:) and definitely barred (B). Galaxies for which no bar
designation can be made are not assigned a rank.

\subsubsection{Disturbance}
\label{disturbance}

We have attempted to classify galaxies on the basis of whether they
appear disturbed or not (Table \ref{gsample}, column 8).  Clearly, this
is important as a possible link to star formation and environment.  We
adopted a $1-4$ rank system which corresponds to the degree of
disturbance (... [no disturbance], D::, D:, D).  In assigning
disturbance class, information was combined from UGC descriptions and
our inspection of direct plate material. Rather uncertain signs of
galaxy disturbance (e.g. slight distortion of outer arms, somewhat
asymmetric appearance) are assigned D:: (rank 2); more definite signs
of distortion (e.g. slight warps, probable tidal plumes, some
disturbance) are assigned D: (rank 3); while significantly disturbed
galaxies (e.g.  bad distortions, strong tidal features, ongoing
merger) are classed as D (rank 4). No rank is given to the few
galaxies for which it is not possible to decide whether a disturbance
is present or not.  An effort was made to keep the disturbance
classification independent of whether there was a nearby companion or
not --- it represented a purely morphological rather than
environmental classification. Of course, there is considerable overlap
between noting a spiral as `peculiar' and as `disturbed', though the
`peculiar' note probably refers to a wider variety of anomalous
morphologies than just disturbance.

\subsubsection{Nearby Companion}
\label{companion}

Finally, while inspecting the galaxies a note was made if there was a
nearby companion (Table \ref{gsample}, column 9).  Although general
limits of $>20\%$ of the size of the galaxy and within $\sim 5$ galaxy
diameters were applied, the `companion' assignment was made on a $1-4$
rank scale (... [no companion], C::, C:, C) depending on the degree of
certainty and/or strength of the interaction.  Smaller galaxies further
away with no signs of distortion are more likely to be projected
companions (C::, rank 2), while larger galaxies closer by with tidal
features are likely to be genuine companions (C, rank 4).

A principal difficulty in defining a robust parameter for the
presence/absence of nearby companion galaxies is the uncertainty
regarding projection effects ---  apparently close galaxies
may be far apart. This is a particular problem in the crowded field of
a cluster.  To help overcome this problem we use two screening
criteria: one using velocity and one using local galaxy surface density. 

First, if velocities were available for both the galaxy and its
companion, and the absolute value of the velocity difference, $|\Delta
v| > 1500$ km s$^{-1}$, then it is assumed either that the projected
companion is a chance superposition or, due to the high velocity
difference, there is negligible tidal interaction. In either case,
these galaxies were no longer considered to have `real' companions and
were grouped with `isolated' galaxies for the subsequent analysis.
These galaxies have their companion parameters listed in square
brackets in Table \ref{gsample}. For Abell 1367 (Paper III, Table 2),
galaxies CGCG nos. 97-125 and 97-133A are also in this category.

In the absence of this velocity criterion, an attempt was made to
estimate the local galaxy surface density. Within an 18 arcmin box
centered on the main galaxy, a count was made of the number of galaxies
of a similar or greater size to the projected companion. For those
cases in which the projected companion was of relatively large size
such that very few, or no similar or larger galaxies were counted in
the 18 arcmin box, the count was repeated for a 1 deg square box.  The
counts were used to estimate the mean surface galaxy density in the
region and the probability, $P$, was computed that the projected
companion was a chance superposition (assuming the galaxies were
distributed randomly across the field).  For $P > 0.05$, the sample
galaxy was omitted from the companion ranking which is given in
parentheses in Table \ref{gsample}.  For Abell 1367 (Paper III, Table
2), galaxies which have been similarly omitted from the companion
ranking are: CGCG nos. 97-044, 97-066, 97-068, 97-120A, 127-036,
127-046, 127-085 and 127-090.

Lastly, for $P \le 0.05$, the projected companion was accepted as a
`real' companion and assigned a rank according to the companion
assignment given in Table \ref{gsample}.

The above procedure which selects galaxies likely to have
tidally-interacting companions is  not quite ideal.  In particular, the
presence of sub-clustering undermines the assumption of random galaxy
distribution around the main galaxy. A cleaner method would require
velocity data for many fainter galaxies which are not yet available.
For the present sample, there are 36 galaxy--companion pairs with $P
\le 0.05$ and known $|\Delta v|$; of these 22\% have $|\Delta v| >
1500$ km s$^{-1}$.  The final selected sample of galaxies with `real'
companions comprises 45 galaxies, of which some 22 have known $|\Delta
v|$. Thus the contamination of the final sample by non
tidally-interacting pairs is expected to be $\ga$ 11\%.

\section {H$\alpha$ Detection and Galaxy Properties}
\label{gprops}

Before investigating the relation between star formation and
environment, it is important to first establish the dependence of star
formation on intrinsic galaxy properties.  This topic has been
discussed in Papers II and III. Making full use of the final galaxy
survey sample, we review here in rather more detail the dependence of
detected H$\alpha$ emission on a variety of galaxy properties and on
the presence or absence of a
nearby companion. We show that the detected H$\alpha$ emission can be
well understood as either normal spiral disk emission, or as
circumnuclear starbursts triggered either by tidal forces on the
galaxy or by a bar.  In at least some cases the tidal forces are due
to a companion galaxy.  The relation of the detected emission (whether
disk emission or circumnuclear starburst) to the cluster environment of
the galaxy will discussed in section \ref{cenviron} below.

\subsection {Apparent and absolute magnitude}
\label{mags}

In Paper III we showed that for galaxies in Abell 1367 the H$\alpha$
detection efficiency was approximately independent of apparent
magnitude down to the CGCG limit, $m_{\rm p}=15.7$.  For our new larger
sample from all 8 clusters (types Sa and later, omitting irregulars or
peculiars) we confirm this earlier result. Kolmogorov-Smirnov (K-S)
tests which compare the cumulative distributions of apparent magnitude
of non-ELGs with either compact ELGs, diffuse ELGs, or all ELGs, all
show no significant differences (significance levels 0.27, 0.71, 0.24
respectively).

Is the same true for absolute magnitude? First we evaluated corrected
magnitudes, $B^{0}_{\rm T}$ following standard methods: converting CGCG
magnitudes, $m_{\rm p}$, first to the $B_{\rm T}$ system following Paturel,
Bottinelli \& Gouguenheim (1994); then correcting 
for galactic and internal absorption following Sandage \&
Tammann (1987).  Finally absolute magnitudes, $M_{\rm B}^{0}$, 
were obtained using
cluster mean redshifts.  Again, K-S tests which compare the cumulative
distributions of absolute magnitude for non-ELGs with either compact
ELGs, diffuse ELGs, or all ELGs, all show no significant differences
(significance levels 0.37, 0.58, 0.60 respectively).

Because H$\alpha$ detection depends on Hubble type (though only
slightly over the range Sa to Sc, see section \ref{htype} below), it
is prudent to ensure that any Hubble type dependence on magnitude is
not confusing these results. A more definitive test, therefore, is to
compare the magnitude distribution of non-ELGs with the magnitude
distribution of an `expected' ELG sample in which each galaxy from the
total sample is weighted by a Hubble type dependent H$\alpha$
detection efficiency. A comparison of this kind also shows no
difference between the distributions of $m_{\rm p}$ and $M_{\rm
B}^{0}$ for the non-ELG and `expected' ELG samples.  We conclude that
star formation rates which render a galaxy detectable in H$\alpha$ are
independent of absolute magnitude in the range $-22 \le M_{\rm B}^{0}
\le -19$ ($H_{0}$ = 75 km ${\rm s}^{-1}$ ${\rm Mpc}^{-1}$) and
independent of apparent magnitude in the range $13 \le m_{\rm p} \le
15.7$ (though they drop below the CGCG limit).
 
These results are broadly consistent with earlier work. For example,
Kennicutt \& Kent (1983) found H$\alpha$ equivalent widths for Sc and
SBc field galaxies to be independent of absolute magnitude in the
range $-22 \le M_{\rm B} \le -17$.  However in more recent work,
Gavazzi, Pierini \& Boselli (1996) find an anti-correlation between
H$\alpha$ equivalent width and galaxy luminosity.  We note that much
of this trend only becomes apparent if low equivalent widths,
$W_{\lambda} \le 10$ \AA\, are included and is not apparent for
equivalent widths restricted to our detection range $W_{\lambda} \ge
20$ \AA.  Furthermore, it is possible that our photographic technique
has missed fainter diffuse H$\alpha$ emission in low luminosity
spirals and this would act to mask the effect noted by Gavazzi et al.

\subsection {Galaxy inclination}
\label{incline}

Does H$\alpha$ detection depend on galaxy inclination?  Axis ratios
for our spiral sample were either taken from Nilson (1973) or measured
from PSS prints (values not given in Table \ref{gsample}).  For spirals
with compact emission, a K-S test shows no significant difference
between the distributions for ELGs and non-ELGs (significance level =
0.28).  For diffuse emission, there is a weak tendency for highly
inclined ($b/a \la 0.3$) galaxies to be less easily detected, but the
effect is only marginal (significance level = 0.06).  These results are
encouraging, partly because galaxy inclination can be ignored in our
subsequent analyses, and partly because both nuclear and disk emission
are unlikely to be masked unless the galaxy is almost edge-on.

\begin{table}
\centering
\caption{\label{tparam} Parameters used in the study}
\begin{tabular}{lcr} \hline
 Categories & Rank/[Bin no.] & $n$ \\ \hline \hline
\multicolumn{3}{l}{\it Emission} \\
 ... & 1 & 180 \\
 S,(S),MS,M,MW,W,(W) & 2 & 115 \\
 && \\
\multicolumn{3}{l}{\it Compact Emission} \\
 ...,D,VD & 1 & 236 \\ 
 VC,C,N,DBL & 2 & 58 \\
 && \\
\multicolumn{3}{l}{\it Diffuse Emission} \\
 ...,VC,C,N,DBL & 1 & 238 \\
 D,VD & 2 & 56 \\ && \\
\multicolumn{3}{l}{\it Type} \\
 Sa & 1,[1] & 62 \\
 Sab & 2,[2] & 30 \\
 Sb & 3,[3] & 49 \\
 Sbc & 4,[4] & 17 \\
 Sc & 5,[5] & 36 \\
 Sc--Irr,Irr & 6,[6] & 18 \\ 
 S  & [7] & 21 \\
 Peculiar  & [8] & 62 \\
 && \\
\multicolumn{3}{l}{\it Bar} \\
 A,A:  & 1 & 37 \\   
 AB,AB: & 2 & 18 \\
 B: & 3 & 21 \\
 B & 4 & 52 \\
 && \\
\multicolumn{3}{l}{\it Disturbed} \\
 ... & 1 & 210 \\
 D:: & 2 & 30 \\
 D: & 3 & 34 \\
 D & 4 & 18 \\
 && \\
\multicolumn{3}{l}{\it Companion} \\
 ...,[C::],[C:],[C] & 1 & 203 \\
 C:: & 2 & 2 \\
 C:  & 3 & 18 \\
 C & 4 & 25 \\ \hline

\end{tabular}
\end{table}

\subsection {Hubble type}
\label{htype}

\begin{figure*}

\epsfxsize=15cm
\hspace*{0.01cm} \epsffile{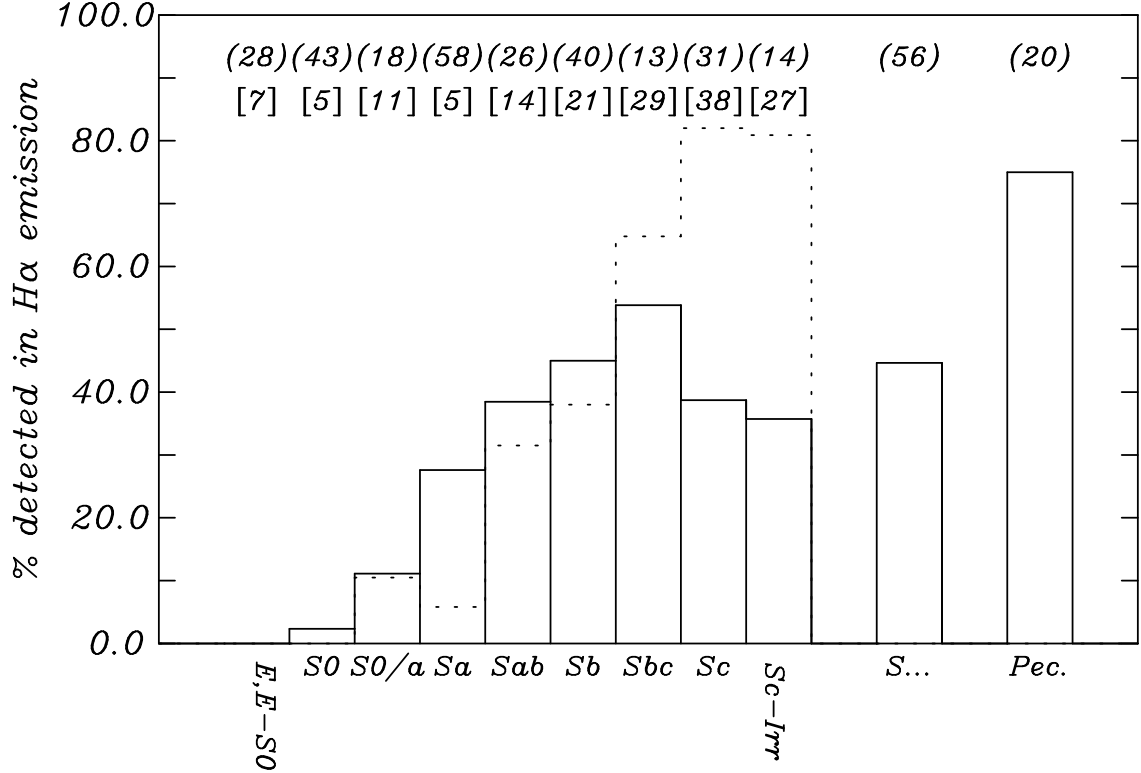}

\caption{\label{etdistrib}
The percentage detection of H$\alpha$+[NII] emission for galaxies
in different Hubble type bins (solid histogram).  
The total sample number for each bin
is given in parentheses.  Also shown are expected percentage detections
for bins in the range E to Sc--Irr (dotted histogram) based on a sample
of field galaxies observed using photoelectric and CCD photometry.
The total field galaxy sample number for each bin is given in square
brackets.
}
\end{figure*}

In Figure \ref{etdistrib} we show the percentage detection of ELGs for
the full range of Hubble types.  Data for early-type galaxies (E, E-S0,
S0 and S0/a) are for Abell 1367 (Paper III) and data for the remaining
types are for all 8 clusters.  The Figure shows a trend of increasing
star formation rate per unit luminosity from early-type to later type
galaxies, which is well-known from previous H$\alpha$, UV and FIR
studies (Kennicutt 1998).

\begin{table*}
\begin{minipage}{165mm}
\caption{\label{gsample} CGCG galaxies surveyed for H$\alpha$ emission}

\begin{tabular}{@{}r@{\hspace{3.0ex}}l@{\hspace{3.0ex}}
r@{\hspace{2.0ex}}r@{\hspace{3.0ex}}r@{\hspace{2.0ex}}
r@{\hspace{2.0ex}}r@{\hspace{3.0ex}}
c@{\hspace{3.0ex}}l@{\hspace{1.0ex}}c@{\hspace{1.0ex}}
c@{\hspace{-1.0ex}}r@{\hspace{3.0ex}}r@{\hspace{3.0ex}}
c@{\hspace{2.0ex}}c@{\hspace{2.0ex}}c}
CGCG & UGC & \multicolumn{4}{c}{R.A.~(1950)~Dec.} &
\multicolumn{1}{c}{$r$~~~} &  $m_{\rm p}$ & Type & 
 Dis. & Cp. & $~~~~v_{\tiny \sun}$ &
 ~~Ref. \hspace{-1.0em} & \multicolumn{2}{c}{H$\alpha$ emission} & 
\hspace{0.5em} Notes \\
 &&&&&& \multicolumn{1}{c}{(r$_{\rm A}$)~~~} 
 &&&&&~~~~(km s$^{-1}$) \hspace{-1.5em} && ~~Vis. & ~~Conc. &  \\ [10pt]
 \multicolumn{2}{l}{\underline{Abell 262}} &&&&&&&&&&&&&& \\
 521-070                   &  1193 &
  1$^{\rm h}$ \hspace{-0.85em} & 
  39\fm5 \hspace{-0.75em}  & 
  +35$^{\circ}$ \hspace{-0.85em} & 
  23$^{\prime}$ \hspace{-0.65em}  & 1.24 &  14.1  & 
 Sab          & ...                   & ...                   &  5114 & 1 & ~~...  & ~~...  & \\
 521-071                   &       &  1 & 40.4 & +36 & 20 & 1.12 &  15.3  &
 SAB:         & ...                   & ...                   & 14723 & 2 & ~~...  & ~~...  & \\
 521-072                   &  1212 &  1 & 41.2 & +34 &  9 & 1.43 &  14.5  &
 SAb          & ...                   & ...                   & 10693 & 1 & ~~...  & ~~...  & \\
 521-073                   &  1220 &  1 & 41.6 & +37 & 26 & 1.29 &  13.6  &
 S pec        & D:: \hspace{-0.900em} & ...                   &  5662 & 1 & ~~S    & ~~D    & \\
 521-074                   &       &  1 & 41.7 & +34 & 25 & 1.28 &  15.0  &
 S: pec       & ...                   & ...                   &  5275 & 2 & ~~S    & ~~VC   & \\
 521-076                   &  1221 &  1 & 41.7 & +37 & 57 & 1.50 &  15.0  &
 Sbc          & ...                   & ...                   & 11095 & 2 & ~~...  & ~~...  & \\
 521-078                   &  1234 &  1 & 42.9 & +34 & 52 & 1.01 &  14.8  &
 Sc/SBc       & D:  \hspace{-0.625em} & ...                   &  5653 & 1 & ~~...  & ~~...  & \\
 521-080                   &  1238 &  1 & 43.4 & +36 & 12 & 0.77 &  13.5  &
 SA:b         & ...                   & ...                   &  4515 & 1 & ~~...  & ~~...  & \\
 521-081                   &       &  1 & 43.5 & +34 & 41 & 1.02 &  14.7  &
 S: pec       & D:  \hspace{-0.625em} & ...                   &  5417 & 1 & ~~...  & ~~...  & \\
 522-003                   &       &  1 & 44.0 & +34 & 32 & 1.04 &  15.2  &
 pec          & D:: \hspace{-0.900em} & ...                   &  4205 & 2 & ~~MS   & ~~N    & \\
 522-004                   &  1248 &  1 & 44.3 & +35 & 18 & 0.74 &  12.9  &
 Sab          & ...                   & ...                   &  4756 & 1 & ~~...  & ~~...  & \\
 522-005                   &  1251 &  1 & 44.6 & +35 & 47 & 0.62 &  15.0  &
 pec          & D                     & C                     &  4845 & 2 & ~~...  & ~~...  & \\
 522-006                   &       &  1 & 44.8 & +34 & 46 & 0.88 &  15.0  &
 SAbc: pec    & ...                   & ...                   &  5557 & 1 & ~~...  & ~~...  & \\
 522-007                   &  1257 &  1 & 45.2 & +36 & 12 & 0.57 &  15.0  &
 SA:ab        & ...                   & ...                   &  4662 & 1 & ~~...  & ~~...  & \\
 522-013                   &       &  1 & 46.5 & +34 & 44 & 0.78 &  15.5  &
 S: pec       & D:: \hspace{-0.900em} & C: \hspace{-0.625em}  &  4025 & 1 & ~~...  & ~~...  & \\
 503-030                   &       &  1 & 47.7 & +33 & 23 & 1.46 &  15.3  &
 S pec        & D                     & ...                   & 15086 & 2 & ~~?    & ~~?    & \\
 522-018                   &  1299 &  1 & 47.7 & +35 &  7 & 0.52 &  15.7  &
 Irr          & ...                   & ...                   &  5498 & 2 & ~~...  & ~~...  & \\
 522-020                   &  1302 &  1 & 47.8 & +35 &  2 & 0.55 &  13.3  &
 SBb          & ...                   & (C::) \hspace{-0.8em} &  4047 & 1 & ~~MS   & ~~C    & \\
 522-021                   &  1307 &  1 & 47.9 & +35 & 40 & 0.27 &  15.1  &
 S            & ...                   & ...                   &  4889 & 1 & ~~MW   & ~~C    & \\
 522-024                   &  1319 &  1 & 48.5 & +35 & 49 & 0.17 &  14.5  &
 SA: pec      & D:: \hspace{-0.900em} & (C:) \hspace{-0.5em}  &  5375 & 1 & ~~W    & ~~VD   & \\
 522-025                   &       &  1 & 49.1 & +35 & 53 & 0.09 &  15.6  &
 SAbc:        & ...                   & ...                   &  6050 & 1 & ~~...  & ~~...  & \\
 522-029A \hspace{-1.08em} &       &  1 & 49.3 & +34 & 55 & 0.57 & (16.4) &
 S            & D:  \hspace{-0.625em} & C                     &       &   & ~~...  & ~~...  & * \\
 522-029B \hspace{-1.08em} &       &  1 & 49.3 & +34 & 55 & 0.57 & (16.4) &
 S            & ...                   & C                     &       &   & ~~...  & ~~...  & * \\
 522-031                   &  1338 &  1 & 49.4 & +35 & 33 & 0.21 &  15.2  &
 SAb          & ...                   & (C::) \hspace{-0.8em} &  4099 & 1 & ~~...  & ~~...  & \\
 522-035                   &  1344 &  1 & 49.7 & +36 & 15 & 0.20 &  14.0  &
 SBa          & ...                   & (C::) \hspace{-0.8em} &  3998 & 1 & ~~?    & ~~?    & \\
 522-038                   &  1347 &  1 & 49.8 & +36 & 22 & 0.27 &  13.9  &
 Sc/SBc       & ...                   & [C:] \hspace{-0.5em}  &  4099 & 1 & ~~...  & ~~...  & \\
 522-041                   &  1349 &  1 & 50.0 & +35 & 48 & 0.06 &  14.3  &
 SABc         & D:: \hspace{-0.900em} & ...                   &  6131 & 1 & ~~W    & ~~VD   & \\
 522-042                   &  1350 &  1 & 50.0 & +36 & 15 & 0.20 &  14.5  &
 SBb          & ...                   & (C::) \hspace{-0.8em} &  5244 & 1 & ~~...  & ~~...  & \\
 503-044                   &       &  1 & 50.1 & +33 & 21 & 1.46 &  15.7  &
 S            & ...                   & ...                   & 11165 & 2 & ~~?    & ~~?    & \\
 522-050                   &  1361 &  1 & 50.9 & +36 & 20 & 0.27 &  15.7  &
 Sc           & ...                   & C:  \hspace{-0.625em} &  5244 & 1 & ~~?    & ~~?    & \\
 522-051                   &       &  1 & 50.9 & +36 & 32 & 0.38 &  15.1  &
 SA           & ...                   &                       &  4686 & 2 & ~~...  & ~~...  & \\
 522-055                   &  1366 &  1 & 51.4 & +36 & 22 & 0.32 &  14.7  &
 SBc          & ...                   & ...                   &  5118 & 1 & ~~...  & ~~...  & \\
 522-058                   &  1385 &  1 & 52.0 & +36 & 41 & 0.51 &  14.2  &
 SBa          & ...                   & C: \hspace{-0.625em}  &  5529 & 1 & ~~S    & ~~N    & \\
 522-059                   &  1380 &  1 & 52.0 & +37 &  5 & 0.72 &  15.6  &
 S            & ...                   & ...                   &  4600 & 1 & ~~...  & ~~...  & \\
 522-060                   &       &  1 & 52.1 & +35 & 11 & 0.48 &  15.1  &
 SBab:        & ...                   & ...                   & 16200 & 1 & ~~...  & ~~...  & \\
 522-062                   &       &  1 & 52.1 & +36 & 41 & 0.51 &  15.2  &
 SBb          & ...                   & (C:) \hspace{-0.5em}  &  5400 & 1 & ~~...  & ~~...  & \\
 522-063                   &  1387 &  1 & 52.2 & +36 &  1 & 0.27 &  15.4  &
 S-Irr        & D:  \hspace{-0.625em} & (C:) \hspace{-0.5em}  &  4502 & 1 & ~~...  & ~~...  & \\
 522-066                   &  1390 &  1 & 52.4 & +36 &  3 & 0.30 &  15.5  &
 S            & ...                   & ...                   &  4368 & 2 & ~~...  & ~~...  & \\
 522-067                   &       &  1 & 52.7 & +37 &  9 & 0.78 &  15.5  &
 Sab: pec:    & D:: \hspace{-0.900em} & ...                   & 14741 & 2 & ~~...  & ~~...  & \\
 522-069                   &  1398 &  1 & 53.0 & +36 & 53 & 0.67 &  14.9  &
 SAc:         & ...                   & ...                   &  5389 & 1 & ~~...  & ~~...  & \\
  522-071                   &  1400 &  1 & 53.2 & +35 & 53 & 0.38 &  13.8  &
 Sb           & ...                   & C:  \hspace{-0.625em} &  4670 & 1 & ~~?    & ~~?    & \\
 522-073                   &  1404 &  1 & 53.4 & +36 & 59 & 0.74 &  15.6  &
 SBb          & ...                   & ...                   &  4458 & 1 & ~~...  & ~~...  & \\
 522-074                   &  1405 &  1 & 53.4 & +37 & 12 & 0.84 &  15.7  &
 Sc           & ...                   & ...                   &  4920 & 1 & ~~...  & ~~...  & \\
 522-075                   &       &  1 & 53.4 & +37 & 15 & 0.87 &  15.7  &
 Irr:         & ...                   & ...                   &  5405 & 1 & ~~...  & ~~...  & \\
 522-077                   &       &  1 & 53.6 & +37 &  5 & 0.80 &  15.5  &
 SBb: pec     & D:: \hspace{-0.900em} & ...                   &  5511 & 2 & ~~M    & ~~N    & \\
 522-078                   &  1411 &  1 & 53.7 & +33 & 56 & 1.21 &  13.9  &
 Sb           & D:: \hspace{-0.900em} & C: \hspace{-0.625em}  &  4748 & 1 & ~~...  & ~~...  & \\
 522-079                   &       &  1 & 53.7 & +35 & 21 & 0.54 &  15.3  &
 SA:c:        & ...                   & ...                   &  5230 & 2 & ~~...  & ~~...  & \\
 522-081                   &  1416 &  1 & 53.8 & +36 & 39 & 0.62 &  14.9  &
 S            & ...                   & ...                   &  5484 & 1 & ~~M    & ~~D    & \\
 522-082                   &       &  1 & 53.9 & +35 & 45 & 0.47 &  15.3  &
 SA:c::       & ...                   & ...                   &  4818 & 1 & ~~?    & ~~?    & \\
 522-086                   &  1437 &  1 & 54.8 & +35 & 40 & 0.58 &  12.6  &
 SAB:c        & D:: \hspace{-0.900em} & (C:) \hspace{-0.5em}  &  4896 & 1 & ~~(S)  & ~~D    & * \\
 522-088                   &  1441 &  1 & 54.9 & +37 &  7 & 0.90 &  15.5  &
 Sb           & ...                   & ...                   &  4996 & 1 & ~~...  & ~~...  & \\
 522-090                   &       &  1 & 55.1 & +34 &  3 & 1.22 &  15.7  &
 S: pec       & ?                     & ...                   & 14279 & 1 & ~~...  & ~~...  & \\
 522-094                   &  1456 &  1 & 56.0 & +36 & 26 & 0.77 &  14.0  &
 Sab          & ...                   & ...                   &  5057 & 1 & ~~...  & ~~...  & \\
 522-095                   &       &  1 & 56.0 & +37 & 30 & 1.15 &  15.6  &
 SAB:b:       & ...                   & ...                   & 14346 & 1 & ~~...  & ~~...  & \\
 522-096                   &  1459 &  1 & 56.1 & +35 & 49 & 0.72 &  15.4  &
 Sc           & ...                   & ...                   &  5466 & 1 & ~~...  & ~~...  & \\
 522-097                   &  1460 &  1 & 56.1 & +36 &  1 & 0.72 &  15.0  &
 Sa pec       & D:: \hspace{-0.900em} & ...                   &  4874 & 1 & ~~...  & ~~...  & \\
 522-100                   &  1474 &  1 & 57.2 & +37 & 21 & 1.18 &  15.0  &
 SB(s)dm      & ...                   & ...                   &  4235 & 1 & ~~MW   & ~~VD   & \\
 522-102                   &  1493 &  1 & 57.9 & +37 & 58 & 1.49 &  14.0  &
 SB:ab        & ...                   & ...                   &  4249 & 1 & ~~W    & ~~D    & \\
  \multicolumn{2}{l}{\underline{Abell 347}} &&&&&&&&&&&&&& \\
 538-034                   &       &  2 & 10.9 & +41 & 39 & 1.45 &  15.0  & 
 S            & D:  \hspace{-0.625em} & ...                   &  4328 & 1 & ~~?    & ~~?    & \\
 538-037                   &  1738 &  2 & 12.8 & +42 & 35 & 1.36 &  15.6  &
 Sc           & ...                   & [C:] \hspace{-0.5em}  &  5734 & 1 & ~~MW   & ~~VD   & \\
 538-038                   &  1743 &  2 & 13.1 & +42 & 35 & 1.33 &  15.7  &
 SBb          & ...                   & [C:] \hspace{-0.5em}  & 13708 & 1 & ~~...  & ~~...  & \\
\end{tabular}    
\end{minipage}    
\end{table*} 

\setcounter{table}{3}

\begin{table*}
\begin{minipage}{165mm}
\caption{\it continued}
\begin{tabular}{@{}r@{\hspace{3.0ex}}l@{\hspace{3.0ex}}
r@{\hspace{2.0ex}}r@{\hspace{3.0ex}}r@{\hspace{2.0ex}}
r@{\hspace{2.0ex}}r@{\hspace{3.0ex}}
c@{\hspace{3.0ex}}l@{\hspace{1.0ex}}c@{\hspace{1.0ex}}
c@{\hspace{-1.0ex}}r@{\hspace{3.0ex}}r@{\hspace{3.0ex}}
c@{\hspace{2.0ex}}c@{\hspace{2.0ex}}c}
CGCG & UGC & \multicolumn{4}{c}{R.A.~(1950)~Dec.} &
\multicolumn{1}{c}{$r$~~~} &  $m_{\rm p}$ & Type & 
 Dis. & Cp. & $~~~~v_{\tiny \sun}$ &
 ~~Ref. \hspace{-1.0em} & \multicolumn{2}{c}{H$\alpha$ emission} & 
\hspace{0.5em} Notes \\
 &&&&&& \multicolumn{1}{c}{(r$_{\rm A}$)~~~} 
 &&&&&~~~~(km s$^{-1}$) \hspace{-1.5em} && ~~Vis. & ~~Conc. &  \\ [10pt]
 538-040                   &  1780 &  
  2$^{\rm h}$ \hspace{-0.85em} & 
  15\fm9 \hspace{-0.75em} & 
  +40$^{\circ}$ \hspace{-0.85em} & 
  20$^{\prime}$ \hspace{-0.65em} & 1.21 &  15.6  &
 Irr          & D:: \hspace{-0.900em} & ...                   &  5204 & 1 & ~~...  & ~~...  & \\
 538-043                   &       &  2 & 16.9 & +41 &  3 & 0.82 &  15.0  &
 pec          & D:  \hspace{-0.625em} & ...                   &  5936 & 3 & ~~S    & ~~DBL  & * \\
 538-045                   &  1796 &  2 & 17.3 & +40 & 34 & 0.98 &  15.5  &
 SAB(s)dm     & ...                   & (C:) \hspace{-0.5em}  &  6983 & 1 & ~~...  & ~~...  & \\
 538-046                   &       &  2 & 17.4 & +41 & 20 & 0.69 &  15.3  &
 SA:b:        & ...                   & ...                   &  5920 & 3 & ~~W    & ~~VD   & \\
 538-047                   &       &  2 & 18.0 & +41 & 35 & 0.58 &  15.6  &
 SB           & ...                   & ...                   &       &   & ~~...  & ~~...  & \\
 538-048                   &       &  2 & 18.2 & +42 & 39 & 0.86 &  15.3  &
 S pec        & D:  \hspace{-0.625em} & C: \hspace{-0.625em}  &  6639 & 1 & ~~MS   & ~~D    & \\
 538-050                   &       &  2 & 19.1 & +42 & 35 & 0.76 &  15.7  &
 Sa:          & ?                     & ...                   &       &   & ~~...  & ~~...  & \\
 538-051                   &  1827 &  2 & 19.1 & +43 & 19 & 1.18 &  15.7  &
 S-Irr        & ...                   & ...                   &  5810 & 2 & ~~...  & ~~...  & \\
 538-052                   &  1831 &  2 & 19.4 & +42 &  7 & 0.51 &  10.8  &
 Sb           & ...                   & ...                   &   527 & 1 & ~~...  & ~~...  & \\
 538-053                   &  1832 &  2 & 19.4 & +42 & 50 & 0.88 &  15.4  &
 Sa           & ...                   & ...                   &  5913 & 1 & ~~...  & ~~...  & \\
 538-054                   &       &  2 & 19.7 & +41 & 56 & 0.41 &  15.7  &
 Sa:          & ...                   & (C::) \hspace{-0.8em} &  6390 & 3 & ~~M    & ~~VD   & \\
 538-056                   &  1840 &  2 & 20.0 & +41 &  9 & 0.47 &  14.1  &
 pec:         & D                     & C                     &  5425 & 2 & ~~...  & ~~...  & * \\
 538-058                   &  1842 &  2 & 20.2 & +41 & 44 & 0.31 &  13.8  &
 Sa           & ...                   & (C::) \hspace{-0.8em} &  5400 & 1 & ~~...  & ~~...  & \\
 538-059                   &       &  2 & 20.8 & +41 & 59 & 0.32 &  15.7  &
 SBb pec      & D:  \hspace{-0.625em} & (C:) \hspace{-0.5em}  &       &   & ~~...  & ~~...  & \\
 538-061                   &  1855 &  2 & 21.4 & +40 & 39 & 0.68 &  15.1  &
 SBa          & ...                   & ...                   & 12849 & 1 & ~~...  & ~~...  & \\
 538-062                   &  1858 &  2 & 21.6 & +41 & 28 & 0.18 &  15.7  &
 SB           & ...                   & ...                   &  5304 & 1 & ~~?    & ~~?    & \\
 538-063                   &       &  2 & 21.6 & +41 & 48 & 0.17 &  15.7  &
 Sbc          & ...                   & ...                   &  5680 & 3 & ~~M    & ~~VD   & \\
 538-066                   &  1866 &  2 & 22.0 & +41 & 38 & 0.09 &  14.9  &
 SBa          & ...                   & (C) \hspace{-0.5em}   &   739 & 1 & ~~...  & ~~...  & \\
 539-014                   &  1868 &  2 & 22.1 & +41 & 52 & 0.16 &  14.4  &
 SBa          & ...                   & (C:) \hspace{-0.5em}  &  4586 & 1 & ~~...  & ~~...  & \\
 539-015                   &       &  2 & 22.2 & +41 & 30 & 0.12 &  15.7  &
 S            & ...                   & (C::) \hspace{-0.8em} &       &   & ~~...  & ~~...  & \\
 539-023                   &  1887 &  2 & 22.9 & +41 & 55 & 0.18 &  13.9  &
 SAc          & ...                   & (C::) \hspace{-0.8em} &  5548 & 1 & ~~...  & ~~...  & \\
 539-024                   &       &  2 & 23.6 & +41 & 37 & 0.11 &  15.0  &
 SBb          & ...                   & ...                   &  5723 & 3 & ~~S    & ~~N    & \\
 539-025                   &       &  2 & 23.7 & +41 & 28 & 0.17 &  15.3  &
 SB pec       & D:  \hspace{-0.625em} & (C:) \hspace{-0.5em}  &  4316 & 3 & ~~S    & ~~N    & \\
 539-026                   &       &  2 & 23.7 & +41 & 48 & 0.16 &  15.7  &
 Sa:          & ...                   & (C:) \hspace{-0.5em}  &  5548 & 1 & ~~...  & ~~...  & \\
 539-027                   &       &  2 & 23.8 & +42 & 35 & 0.63 &  15.7  &
 SB:bc:       & ...                   & ...                   &       &   & ~~...  & ~~...  & \\
 539-029                   &       &  2 & 24.3 & +41 & 42 & 0.20 &  15.7  &
 S            & D:  \hspace{-0.625em} & (C:) \hspace{-0.5em}  &  6740 & 3 & ~~MS   & ~~VD   & \\
 539-030                   &  1915 &  2 & 24.4 & +41 & 45 & 0.22 &  14.4  &
 Sb:          & D:: \hspace{-0.900em} & (C) \hspace{-0.5em}   &  5638 & 3 & ~~S    & ~~D    & \\
 539-032                   &  1961 &  2 & 26.3 & +42 &  2 & 0.51 &  15.0  &
 SB:c         & ...                   & ...                   &  5631 & 1 & ~~...  & ~~...  & \\
 539-036                   &  1988 &  2 & 28.1 & +40 & 10 & 1.19 &  14.7  &
 Sab          & ...                   & ...                   &  5814 & 1 & ~~S    & ~~D    & \\
 539-038                   &       &  2 & 28.3 & +40 &  2 & 1.27 &  15.7  &
 S pec        & D:: \hspace{-0.900em} & C                     &  5889 & 3 & ~~S    & ~~N    & \\
 539-040                   &  1997 &  2 & 29.0 & +43 & 14 & 1.29 &  15.4  &
 Sb           & ...                   & ...                   &  6162 & 1 & ~~...  & ~~...  & \\
 539-041                   &  2001 &  2 & 29.2 & +41 & 59 & 0.83 &  14.6  &
 Sab          & ...                   & (C::) \hspace{-0.8em} &  6989 & 1 & ~~...  & ~~...  & \\
 539-046                   &  2034 &  2 & 30.6 & +40 & 19 & 1.32 &  15.0  &
 Irr          & ...                   & [C::] \hspace{-0.8em} &   579 & 2 & ~~...  & ~~...  & \\
 539-048                   &       &  2 & 30.8 & +42 & 28 & 1.13 &  15.7  &
 S            & ...                   & ...                   &       &   & ~~...  & ~~...  & \\
 539-052                   &  2058 &  2 & 31.8 & +40 & 55 & 1.23 &  15.6  &
 Sb/SBc       & ...                   & ...                   &       &   & ~~...  & ~~...  & \\
 539-053                   &  2060 &  2 & 31.9 & +41 &  9 & 1.18 &  14.7  &
 SBab         & ...                   & ...                   &  4581 & 1 & ~~...  & ~~...  & \\
 539-056                   &  2066 &  2 & 32.3 & +40 & 40 & 1.36 &  13.2  &
 Sa           & ...                   & C: \hspace{-0.625em}  &  5843 & 1 & ~~...  & ~~...  & \\
   \multicolumn{2}{l}{\underline{Abell 400}} &&&&&&&&&&&&&& \\ 
 415-021                   &  2372 &  2 & 51.3 & + 5 & 47 & 0.77 &  15.5  &
 SAB:c        & ...                   & ...                   &  7910 & 1 & ~~W    & ~~D    & \\
 415-022                   &  2375 &  2 & 51.4 & + 6 &  3 & 0.77 &  15.5  &
 S            & ...                   & ...                   &  7607 & 1 & ~~...  & ~~...  & \\
 415-025                   &       &  2 & 52.7 & + 5 & 55 & 0.48 &  15.7  &
 S            & ...                   & (C:) \hspace{-0.5em}  &  7453 & 1 & ~~W    & ~~VD   & \\
 415-027                   &       &  2 & 53.1 & + 6 &  8 & 0.47 &  15.6  &
 S            & ...                   & ...                   &  6760 & 1 & ~~...  & ~~...  & \\
 415-028                   &  2399 &  2 & 53.2 & + 6 &  0 & 0.40 &  15.3  &
 SAB:c        & ...                   & ...                   &  8006 & 1 & ~~...  & ~~...  & \\
 415-030                   &  2405 &  2 & 53.3 & + 6 & 17 & 0.51 &  15.1  &
 Sc           & ...                   & ...                   &  7709 & 1 & ~~W    & ~~D    & \\
 415-031                   &  2414 &  2 & 53.7 & + 4 & 20 & 1.28 &  15.5  &
 Sc           & ...                   & C:  \hspace{-0.625em} &  8267 & 1 & ~~...  & ~~...  & \\
 415-032                   &  2415 &  2 & 53.7 & + 5 & 57 & 0.29 &  15.5  &
 SBbc         & ...                   & [C] \hspace{-0.5em}   &  6590 & 1 & ~~...  & ~~...  & \\
 415-035                   &  2419 &  2 & 54.0 & + 7 &  8 & 1.10 &  14.8  &
 SBa          & ...                   & ...                   &  8090 & 1 & ~~...  & ~~...  & \\
 415-037                   &  2423 &  2 & 54.1 & + 4 & 47 & 0.89 &  15.7  &
 Sc           & ...                   & ...                   &  7724 & 1 & ~~...  & ~~...  & \\
 415-039                   &  2426 &  2 & 54.5 & + 5 &  7 & 0.61 &  15.1  &
 SA:b         & ...                   & ...                   &  7460 & 1 & ~~...  & ~~...  & \\
 415-042                   &       &  2 & 55.1 & + 5 & 45 & 0.07 &  15.7  &
 S: pec       & D:  \hspace{-0.625em} & (C:) \hspace{-0.5em}  &  7200 & 2 & ~~...  & ~~...  & \\
 415-048                   &  2444 &  2 & 55.8 & + 6 &  6 & 0.28 &  15.2  &
 S            & ...                   & ...                   &  6708 & 1 & ~~M    & ~~D    & \\
 415-053                   &  2469 &  2 & 57.7 & + 5 & 31 & 0.62 &  15.2  &
 pec:         & ...                   & (C:) \hspace{-0.5em}  &  8617 & 1 & ~~...  & ~~...  & \\
 415-058                   &       &  3 &  2.0 & + 5 & 15 & 1.53 &  15.7  &
 Sbc:         & ...                   & ...                   &  8312 & 2 & ~~(W)  & ~~VD   & \\
 \multicolumn{2}{l}{\underline{Abell 426}} &&&&&&&&&&&&&& \\
 540-036                   &       &  3 &  2.9 & +41 & 33 & 1.46 &  15.7  &
 S:c: pec     & D:  \hspace{-0.625em} & ...                   &  3610 & 2 & ~~?    & ~~?    & \\
 540-039                   &  2534 &  3 &  3.3 & +41 & 17 & 1.41 &  15.7  &
 pec:         & ...                   & ...                   &  5306 & 1 & ~~...  & ~~...  & \\
 540-042                   &  2538 &  3 &  3.8 & +41 & 34 & 1.35 &  15.6  & 
 SBa          & ...                   & ...                   &  4046 & 1 & ~~?    & ~~?    & \\
 540-043                   &  2544 &  3 &  4.2 & +42 & 12 & 1.40 &  15.0  &
 S            & ...                   & ...                   &  5198 & 1 & ~~...  & ~~...  & \\
 540-047                   &  2561 &  3 &  6.4 & +40 & 48 & 1.10 &  15.5  &
 Sb           & ...                   & ...                   &  5821 & 1 & ~~...  & ~~...  & \\
 540-049                   &  2567 &  3 &  7.0 & +40 & 35 & 1.09 &  14.3  &
 S-Irr        & ...                   & ...                   &  3018 & 1 & ~~M    & ~~VD   & \\
 540-058                   &       &  3 &  9.9 & +42 & 49 & 1.12 &  15.7  &
 Sb pec       & D:: \hspace{-0.900em} & ...                   &  9011 & 2 & ~~...  & ~~...  & \\
 525-009                   &  2604 &  3 & 11.5 & +39 & 27 & 1.26 &  14.8  &
 SBc          & ...                   & ...                   &  4520 & 1 & ~~...  & ~~...  & \\
 540-064                   &  2608 &  3 & 11.7 & +41 & 51 & 0.53 &  14.0  &
 SBb          & D:  \hspace{-0.625em} & C:  \hspace{-0.625em} &  7042 & 1 & ~~S    & ~~N    & \\
 525-011                   &  2610 &  3 & 11.8 & +39 & 11 & 1.41 &  15.7  &
 Sb           & ...                   & ...                   &  5090 & 1 & ~~...  & ~~...  & \\
\end{tabular}    
\end{minipage}    
\end{table*} 

\setcounter{table}{3}

\begin{table*}
\begin{minipage}{165mm}
\caption{\it continued}
\begin{tabular}{@{}r@{\hspace{3.0ex}}l@{\hspace{3.0ex}}
r@{\hspace{2.0ex}}r@{\hspace{3.0ex}}r@{\hspace{2.0ex}}
r@{\hspace{2.0ex}}r@{\hspace{3.0ex}}
c@{\hspace{3.0ex}}l@{\hspace{1.0ex}}c@{\hspace{1.0ex}}
c@{\hspace{-1.0ex}}r@{\hspace{3.0ex}}r@{\hspace{3.0ex}}
c@{\hspace{2.0ex}}c@{\hspace{2.0ex}}c}
CGCG & UGC & \multicolumn{4}{c}{R.A.~(1950)~Dec.} &
\multicolumn{1}{c}{$r$~~~} &  $m_{\rm p}$ & Type & 
 Dis. & Cp. & $~~~~v_{\tiny \sun}$ &
 ~~Ref. \hspace{-1.0em} & \multicolumn{2}{c}{H$\alpha$ emission} & 
\hspace{0.5em} Notes \\
 &&&&&& \multicolumn{1}{c}{(r$_{\rm A}$)~~~} 
 &&&&&~~~~(km s$^{-1}$) \hspace{-1.5em} && ~~Vis. & ~~Conc. &  \\ [10pt]
 540-065                   &  2612 &  
 3$^{\rm h}$ \hspace{-0.85em} & 
 11\fm9 \hspace{-0.75em} & 
 +41$^{\circ}$ \hspace{-0.85em} & 
 48$^{\prime}$ \hspace{-0.65em}  & 0.49 &  15.4  &
 Sc           & ...                   & ...                   &  6446 & 1 & ~~?    & ~~?    & \\
 540-067                   &       &  3 & 12.0 & +41 & 25 & 0.39 &  15.3  &
 SA:a:        & ...                   & ...                   &  5945 & 1 & ~~M    & ~~D    & \\
 540-069                   &  2617 &  3 & 12.7 & +40 & 43 & 0.49 &  14.3  &
 SABc         & D:: \hspace{-0.900em} & (C:) \hspace{-0.5em}  &  4627 & 1 & ~~M    & ~~C    & \\
 540-070                   &  2618 &  3 & 12.7 & +41 & 53 & 0.46 &  14.9  &
 Sab          & ...                   & ...                   &  5376 & 1 & ~~W    & ~~D    & \\
 540-071                   &       &  3 & 12.7 & +42 & 44 & 0.93 &  15.6  &
 SA:a:        & ...                   & ...                   &       &   & ~~MS   & ~~N    & \\
 540-073                   &  2621 &  3 & 13.2 & +41 & 21 & 0.25 &  14.7  &
 Sa           & ...                   & ...                   &  4747 & 1 & ~~...  & ~~...  & \\
 540-076                   &  2625 &  3 & 13.5 & +39 & 50 & 0.96 &  15.7  &
 S: pec:      & ...                   & ...                   &  4252 & 1 & ~~...  & ~~...  & \\
 540-078                   &  2626 &  3 & 13.7 & +41 & 10 & 0.21 &  15.7  &
 Sa:          & ...                   & ...                   &  6418 & 2 & ~~...  & ~~...  & \\
 540-083                   &  2639 &  3 & 14.5 & +41 & 47 & 0.30 &  15.6  &
 Sab          & ...                   & ...                   &  4046 & 1 & ~~...  & ~~...  & \\
 540-084                   &  2640 &  3 & 14.5 & +43 &  7 & 1.12 &  14.8  &
 SBb          & ...                   & ...                   &  6161 & 1 & ~~MS   & ~~D    & \\
 540-090                   &  2654 &  3 & 15.4 & +42 &  7 & 0.49 &  14.6  &
 pec:         & D:  \hspace{-0.625em} & C:  \hspace{-0.625em} &  5793 & 1 & ~~...  & ~~...  & \\
 540-091                   &  2655 &  3 & 15.4 & +43 &  3 & 1.07 &  14.1  &
 SBc          & ...                   & ...                   &  6155 & 1 & ~~M    & ~~C    & * \\
 540-093                   &  2658 &  3 & 15.5 & +41 & 18 & 0.03 &  14.5  &
 SAb          & D:  \hspace{-0.625em} & (C:) \hspace{-0.5em}  &  3124 & 1 & ~~...  & ~~...  & \\
 540-094                   &  2659 &  3 & 15.6 & +40 & 25 & 0.57 &  14.9  &
 Sbc          & ...                   & ...                   &  6193 & 1 & ~~S    & ~~N    & \\
 540-100                   &  2665 &  3 & 16.2 & +41 & 27 & 0.13 &  15.5  &
 Sc? pec      & D:: \hspace{-0.900em} & (C:) \hspace{-0.5em}  &  7861 & 1 & ~~MW   & ~~VD   & \\
 540-103                   &  2669 &  3 & 16.5 & +41 & 20 & 0.14 &  13.0  &
 pec:         & ...                   & ...                   &  5264 & 1 & ~~S    & ~~N    & \\
 540-106                   &  2672 &  3 & 16.8 & +40 & 44 & 0.41 &  15.7  &
 Sa?          & ...                   & ...                   &  4295 & 1 & ~~...  & ~~...  & \\
 525-021                   &       &  3 & 17.0 & +39 & 23 & 1.24 &  15.5  &
 SBa:         & ...                   & ...                   &       &   & ~~...  & ~~...  & \\
 540-112A \hspace{-1.08em} &  2688 &  3 & 18.0 & +41 & 45 & 0.41 & (15.4) &
 pec          & D:  \hspace{-0.625em} & C                     &  3015 & 1 & ~~MS   & ~~C    & * \\
 540-112B \hspace{-1.08em} &  2688 &  3 & 18.0 & +41 & 45 & 0.41 & (16.2) &
 S: pec       & D                     & C                     &  2882 & 4 & ~~?    & ~~?    & * \\
 540-114                   &       &  3 & 18.3 & +40 & 15 & 0.76 &  15.6  &
 S:a:         & ...                   & ...                   &       &   & ~~...  & ~~...  & \\
 540-115                   &       &  3 & 18.3 & +41 & 19 & 0.35 &  15.6  &
 Sa:          & ...                   & (C:) \hspace{-0.5em}  &  3343 & 1 & ~~?    & ~~?    & \\
 540-118                   &  2696 &  3 & 18.7 & +42 &  0 & 0.57 &  15.7  &
 S:           & ...                   & ...                   &  5454 & 1 & ~~...  & ~~...  & \\
 540-121                   &  2700 &  3 & 19.6 & +42 & 22 & 0.82 &  15.5  &
 SB:b         & ...                   & ...                   &  6622 & 1 & ~~MW   & ~~N    & \\
 541-003                   &       &  3 & 22.2 & +40 & 21 & 1.02 &  14.9  &
 SAa:         & ?                     & ...                   &       &   & ~~M    & ~~D    & \\
 541-005                   &  2730 &  3 & 22.6 & +40 & 35 & 0.98 &  15.3  &
 Sb           & ...                   & [C:] \hspace{-0.5em}  &  3772 & 2 & ~~...  & ~~...  & \\
 541-006                   &  2732 &  3 & 22.8 & +40 & 37 & 0.99 &  15.4  &
 SBb          & ...                   & [C:] \hspace{-0.5em}  &  6966 & 1 & ~~...  & ~~...  & \\
 541-008                   &  2736 &  3 & 23.2 & +40 & 20 & 1.12 &  14.7  &
 Sab          & ...                   & ...                   &  5887 & 1 & ~~...  & ~~...  & \\
 541-009                   &  2742 &  3 & 24.4 & +40 & 44 & 1.14 &  15.5  &
 SBc          & ...                   & ...                   &  4401 & 1 & ~~M    & ~~C    & \\
 541-011                   &       &  3 & 25.2 & +39 & 59 & 1.44 &  15.0  &
 SB:b: pec    & D:  \hspace{-0.625em} & (C:)\hspace{-0.625em} &  4246 & 1 & ~~S    & ~~N    & \\
 541-017                   &  2759 &  3 & 26.7 & +41 & 40 & 1.35 &  14.8  &
 pec:         & D:  \hspace{-0.625em} & ...                   &  4237 & 1 & ~~S    & ~~D    & \\
 \multicolumn{2}{l}{\underline{Abell 569}} &&&&&&&&&&&&&& \\ 
 234-043                   &  3638 &  6 & 59.2 & +49 & 30 & 0.88 &  14.4  &
 SB:ab        & ...                   & ...                   &  5567 & 1 & ~~MW   & ~~VD   & \\
 234-050                   &  3662 &  7 &  2.1 & +50 & 35 & 1.34 &  14.6  &
 SBa:         & ...                   & (C::) \hspace{-0.8em} &  6276 & 1 & ~~...  & ~~...  & \\
 234-051                   &  3663 &  7 &  2.1 & +50 & 50 & 1.50 &  14.8  &
 SBa          & ...                   & ...                   &  6290 & 1 & ~~...  & ~~...  & \\
 234-055                   &       &  7 &  3.5 & +48 & 25 & 0.29 &  15.6  &
 S            & ...                   & ...                   &  5882 & 1 & ~~...  & ~~...  & \\
 234-056                   &       &  7 &  3.8 & +48 & 58 & 0.26 &  14.8  &
 S pec        & ...                   & C:  \hspace{-0.625em} &  6212 & 2 & ~~S    & ~~N    & \\
 234-057                   &       &  7 &  4.0 & +48 & 29 & 0.22 &  15.7  &
 pec          & ...                   & ...                   &       &   & ~~M    & ~~N    & \\
 234-060                   &  3681 &  7 &  4.3 & +50 & 45 & 1.40 &  14.3  &
 SBb          & ...                   & (C:) \hspace{-0.5em}  &  5985 & 1 & ~~...  & ~~...  & \\
 234-061                   &       &  7 &  4.4 & +49 &  0 & 0.23 &  15.5  &
 SAa:         & ...                   & (C::) \hspace{-0.8em} &  6236 & 1 & ~~W    & ~~VD   & \\
 234-062                   &       &  7 &  4.4 & +49 & 13 & 0.37 &  15.1  &
 SB:a:        & ...                   & (C::) \hspace{-0.8em} &  5860 & 1 & ~~...  & ~~...  & \\
 234-065                   &       &  7 &  4.7 & +48 & 12 & 0.35 &  15.6  &
 SB: pec      & D:  \hspace{-0.625em} & (C::) \hspace{-0.8em} &       &   & ~~MW   & ~~VD   & \\
 234-066                   &  3687 &  7 &  4.7 & +50 & 42 & 1.37 &  15.5  &
 pec:         & D:: \hspace{-0.900em} & (C:) \hspace{-0.5em}  &  6164 & 2 & ~~M    & ~~N    & \\
 234-067                   &       &  7 &  5.0 & +49 &  4 & 0.25 &  15.1  &
 Sa:          & ...                   & ...                   &  6258 & 1 & ~~M    & ~~D    & \\
 234-069                   &       &  7 &  5.3 & +48 & 39 & 0.04 &  15.6  &
 Sa:          & ...                   & C:  \hspace{-0.625em} &  5296 & 2 & ~~W    & ~~D    & \\
 234-071                   &       &  7 &  5.4 & +49 & 54 & 0.82 &  15.5  &
 SB: pec      & D:  \hspace{-0.625em} & (C::) \hspace{-0.8em} &  4662 & 1 & ~~S    & ~~C    & \\
 234-079A \hspace{-1.08em} &  3706 &  7 &  6.1 & +47 & 59 & 0.50 & (15.3) &
 S: pec       & D                     & C                     &  6115 & 2 & ~~MS   & ~~N    & * \\
 234-079B \hspace{-1.08em} &  3706 &  7 &  6.1 & +47 & 59 & 0.50 & (15.7) &
 S: pec       & D                     & C                     &  6077 & 2 & ~~...  & ~~...  & * \\
 234-088A \hspace{-1.08em} &  3719 &  7 &  7.2 & +48 & 35 & 0.22 & (15.4) &
 Sab          & D:: \hspace{-0.900em} & C                     &  5820 & 2 & ~~...  & ~~...  & * \\
 234-090                   &       &  7 &  7.2 & +49 &  5 & 0.33 &  15.2  &
 Sbc          & ...                   & ...                   &  5956 & 1 & ~~M    & ~~VD   & \\
 234-092                   &       &  7 &  7.3 & +49 & 58 & 0.89 &  15.7  &
 Sa:          & ...                   & ...                   &  6296 & 2 & ~~...  & ~~...  & \\
 234-093                   &  3724 &  7 &  7.7 & +48 & 19 & 0.37 &  14.5  &
 SBb          & ...                   & ...                   &  5925 & 1 & ~~MW   & ~~D    & \\
 234-094                   &       &  7 &  7.7 & +49 & 10 & 0.41 &  15.4  & 
 S-Irr        & ...                   & ...                   &  6089 & 2 & ~~W    & ~~VD   & \\
 234-100                   &  3734 &  7 &  8.7 & +47 & 15 & 1.06 &  13.2  &
 SAb          & ...                   & ...                   &   955 & 1 & ~~...  & ~~...  & \\
 234-102                   &       &  7 &  9.1 & +49 &  5 & 0.49 &  15.0  &
 Sb:          & D:: \hspace{-0.900em} & (C:) \hspace{-0.5em}  &       &   & ~~...  & ~~...  & \\
 234-103                   &       &  7 &  9.1 & +49 & 51 & 0.89 &  15.2  &
 Sa:          & ...                   & ...                   &       &   & ~~...  & ~~...  & \\
 234-107                   &  3741 &  7 & 10.0 & +50 & 20 & 1.22 &  15.5  &
 Sc           & ...                   & ...                   &  5301 & 1 & ~~...  & ~~...  & \\
 234-114                   &       &  7 & 12.5 & +48 & 21 & 0.84 &  15.6  &
 SAa:         & ...                   & ...                   &       &   & ~~...  & ~~...  & \\
 235-005                   &       &  7 & 15.1 & +49 & 18 & 1.16 &  15.5  &
 SA           & ...                   & ...                   &       &   & ~~W    & ~~VD   & \\
 235-007                   &       &  7 & 16.8 & +49 & 11 & 1.32 &  15.0  &
 Sbc:         & ...                   & ...                   &       &   & ~~MS   & ~~VD   & * \\
 \multicolumn{2}{l}{\underline{Abell 779}} &&&&&&&&&&&&&& \\
 180-057                   &  4843 &  9 &  9.6 & +35 &  7 & 1.49 &  14.2  &
 SB           & ...                   & ...                   &  1951 & 1 & ~~...  & ~~...  & \\
 180-059                   &       &  9 & 10.6 & +33 & 31 & 1.10 &  15.7  &
 S            & ...                   & ...                   &  3393 & 1 & ~~...  & ~~...  & \\
 180-060                   &       &  9 & 10.6 & +35 &  2 & 1.32 &  15.4  &
 Sa:          & ...                   & ...                   &  7200 & 1 & ~~MW   & ~~D    & \\
\end{tabular}    
\end{minipage}    
\end{table*} 

\setcounter{table}{3}

\begin{table*}
\begin{minipage}{165mm}
\caption{\it continued}
\begin{tabular}{@{}r@{\hspace{3.0ex}}l@{\hspace{3.0ex}}
r@{\hspace{2.0ex}}r@{\hspace{3.0ex}}r@{\hspace{2.0ex}}
r@{\hspace{2.0ex}}r@{\hspace{3.0ex}}
c@{\hspace{3.0ex}}l@{\hspace{1.0ex}}c@{\hspace{1.0ex}}
c@{\hspace{-1.0ex}}r@{\hspace{3.0ex}}r@{\hspace{3.0ex}}
c@{\hspace{2.0ex}}c@{\hspace{2.0ex}}c}
CGCG & UGC & \multicolumn{4}{c}{R.A.~(1950)~Dec.} &
\multicolumn{1}{c}{$r$~~~} &  $m_{\rm p}$ & Type & 
 Dis. & Cp. & $~~~~v_{\tiny \sun}$ &
 ~~Ref. \hspace{-1.0em} & \multicolumn{2}{c}{H$\alpha$ emission} & 
\hspace{0.5em} Notes \\
 &&&&&& \multicolumn{1}{c}{(r$_{\rm A}$)~~~} 
 &&&&&~~~~(km s$^{-1}$) \hspace{-1.5em} && ~~Vis. & ~~Conc. &  \\ [10pt]
 181-006                   &  4894 &  
 9$^{\rm h}$ \hspace{-0.85em} & 
 13\fm7 \hspace{-0.75em} & 
 +34$^{\circ}$ \hspace{-0.85em} & 
 39$^{\prime}$ \hspace{-0.65em} & 0.74 &  13.9  &
 SB pec       & D:  \hspace{-0.625em} & C: \hspace{-0.625em}  &  1681 & 1 & ~~MW   & ~~N    & * \\
 151-048                   &  4908 &  9 & 14.2 & +32 & 13 & 1.48 &  15.7  &
 Sb           & ...                   & ...                   & 14737 & 1 & ~~...  & ~~...  & \\
 181-007                   &       &  9 & 14.9 & +34 & 43 & 0.67 &  15.7  &
 SA:a:        & ...                   & ...                   &  7002 & 2 & ~~...  & ~~...  & \\
 151-053                   &       &  9 & 15.5 & +32 & 28 & 1.23 &  15.6  &
 SB:          & ...                   & ...                   &  8042 & 1 & ~~...  & ~~...  & \\
 181-012                   &       &  9 & 15.5 & +34 & 30 & 0.47 &  15.5  &
 Sa:          & ...                   & ...                   &  7198 & 1 & ~~...  & ~~...  & \\
 181-013                   &  4926 &  9 & 15.5 & +34 & 46 & 0.66 &  15.4  &
 Sb:          & ...                   & C: \hspace{-0.625em}  &  6365 & 1 & ~~W    & ~~VD   & \\
 181-016                   &  4935 &  9 & 16.2 & +34 & 13 & 0.21 &  15.7  &
 SBa          & ...                   & ...                   &  6960 & 1 & ~~...  & ~~...  & \\
 181-017                   &       &  9 & 16.3 & +33 & 57 & 0.09 &  15.3  &
 S:a:         & ...                   & (C:) \hspace{-0.5em}  &  6106 & 1 & ~~...  & ~~...  & \\
 181-019                   &       &  9 & 16.4 & +34 & 31 & 0.43 &  15.6  &
 pec          & D:: \hspace{-0.900em} & C: \hspace{-0.625em}  & 13783 & 2 & ~~...  & ~~...  & \\
 181-023                   &  4941 &  9 & 16.7 & +33 & 57 & 0.03 &  15.4  &
 S            & D:  \hspace{-0.625em} & C                     &  6106 & 1 & ~~W    & ~~N    & \\
 181-026                   &  4947 &  9 & 16.9 & +33 &  8 & 0.68 &  15.3  &
 SB           & D:  \hspace{-0.625em} & ...                   & 13790 & 1 & ~~?    & ~~?    & \\
 181-030                   &       &  9 & 17.6 & +33 & 17 & 0.58 &  15.5  &
 SB:b         & ...                   & ...                   &  6449 & 1 & ~~MW   & ~~N    & \\
 181-032                   &  4960 &  9 & 17.8 & +35 & 35 & 1.29 &  14.8  &
 SBb          & ...                   & ...                   &  7544 & 1 & ~~MW   & ~~D    & \\
 181-036                   &       &  9 & 19.2 & +34 &  8 & 0.42 &  15.7  &
 S            & ...                   & (C::) \hspace{-0.8em} &  6025 & 1 & ~~...  & ~~...  & \\
 181-037                   &  4988 &  9 & 20.2 & +34 & 56 & 0.94 &  15.7  &
 SABm         & ...                   & ...                   &  1575 & 1 & ~~...  & ~~...  & \\
 181-042                   &       &  9 & 21.9 & +33 & 57 & 0.85 &  15.6  &
 SBbc:        & ...                   & C:: \hspace{-0.900em} & 12679 & 2 & ~~...  & ~~...  & \\
 181-043                   &  5015 &  9 & 22.7 & +34 & 30 & 1.06 &  15.7  &
 SABdm        & ...                   & ...                   &  1646 & 1 & ~~...  & ~~...  & \\
 181-044                   &  5020 &  9 & 23.0 & +34 & 52 & 1.24 &  15.3  &
 Sc           & ...                   & C                     &  1630 & 1 & ~~...  & ~~...  & \\
 181-045                   &       &  9 & 24.2 & +34 & 39 & 1.33 &  15.7  &
 S:b:         & ...                   & ...                   &  6465 & 1 & ~~...  & ~~...  & \\
 \multicolumn{2}{l}{\underline{Abell 1656}} &&&&&&&&&&&&&& \\
 159-109                   &  8024 & 12 & 51.6 & +27 & 25 & 1.24 &  14.9  &
 Irr          & ...                   & ...                   &   376 & 2 & ~~...  & ~~...  & \\
 159-116                   &  8033 & 12 & 52.2 & +29 & 12 & 1.20 &  12.3  &
 Sc           & ...                   & ...                   &  2453 & 1 & ~~S    & ~~N    & * \\
 160-025                   &  8060 & 12 & 54.1 & +27 & 15 & 1.00 &  14.0  &
 SBa          & ...                   & (C) \hspace{-0.5em}   &  6404 & 1 & ~~...  & ~~...  & \\
 160-038                   &  8069 & 12 & 54.8 & +29 & 18 & 0.97 &  14.8  &
 SB:          & D:: \hspace{-0.900em} & (C::) \hspace{-0.8em} &  7472 & 1 & ~~?    & ~~?    & \\
 160-043                   &  8071 & 12 & 55.1 & +28 & 28 & 0.45 &  15.4  &
 S            & ...                   & C                     &  7069 & 1 & ~~...  & ~~...  & \\
 160-050                   &  8076 & 12 & 55.4 & +29 & 55 & 1.40 &  15.2  &
 SAB:c        & ...                   & ...                   &  5304 & 1 & ~~MW   & ~~VD   & \\
 160-055                   &  8082 & 12 & 55.7 & +28 & 31 & 0.37 &  14.2  &
 SB:ab        & D:: \hspace{-0.900em} & ...                   &  7227 & 1 & ~~S    & ~~N    & * \\
 160-058                   &       & 12 & 55.8 & +28 & 59 & 0.66 &  15.5  &
 S            & ...                   & ...                   &  7609 & 1 & ~~M    & ~~D    & \\
 160-062A \hspace{-1.08em} &       & 12 & 55.9 & +29 & 24 & 0.97 & (15.8) &
 pec          & D                     & C                     &  7837 & 2 & ~~...  & ~~...  & * \\
 160-062B \hspace{-1.08em} &       & 12 & 55.9 & +29 & 24 & 0.97 & (15.8) &
 pec          & D                     & C                     &       &   & ~~...  & ~~...  & * \\
 160-064                   &       & 12 & 56.1 & +27 & 31 & 0.64 &  15.4  &
 pec          & D:  \hspace{-0.625em} & ...                   &  7368 & 1 & ~~S    & ~~N    & \\
 160-067                   &       & 12 & 56.2 & +27 & 26 & 0.70 &  15.4  &
 pec          & D:: \hspace{-0.900em} & ...                   &  7664 & 1 & ~~S    & ~~N    & \\
 160-073                   &  8096 & 12 & 56.5 & +28 &  6 & 0.20 &  14.9  &
 S            & ...                   & ...                   &  7526 & 1 & ~~...  & ~~...  & \\
 160-075                   &       & 12 & 56.6 & +28 & 23 & 0.18 &  15.5  &
 pec          & D:  \hspace{-0.625em} & [C] \hspace{-0.5em}   &  9386 & 1 & ~~M    & ~~N    & \\
 160-099                   &       & 12 & 57.2 & +28 & 54 & 0.53 &  15.6  &
 Sa:          & ...                   & ...                   &  5327 & 1 & ~~MS   & ~~N    & \\
 160-110                   &  8108 & 12 & 57.6 & +27 & 10 & 0.88 &  14.7  &
 S            & ...                   & ...                   &  5898 & 1 & ~~...  & ~~...  & \\
 160-113A \hspace{-1.08em} &       & 12 & 57.7 & +28 &  8 & 0.11 & (16.0) &
 pec          & ...                   & [C] \hspace{-0.5em}   &  5128 & 2 & ~~MW   & ~~N    & * \\
 160-127                   &       & 12 & 58.1 & +27 & 55 & 0.30 &  15.4  &
 pec          & D:: \hspace{-0.900em} & ...                   &  7476 & 1 & ~~S    & ~~N    & \\
 160-130                   &       & 12 & 58.2 & +28 & 20 & 0.16 &  15.1  &
 pec:         & D:: \hspace{-0.900em} & ...                   &  7633 & 1 & ~~S    & ~~N    & \\
 160-132                   &  8118 & 12 & 58.2 & +29 & 17 & 0.85 &  14.6  &
 S            & ...                   & ...                   &  7275 & 1 & ~~...  & ~~...  & \\
 160-139                   &       & 12 & 58.4 & +28 & 26 & 0.23 &  14.6  &
 SB:ab        & ...                   & ...                   &  5807 & 1 & ~~...  & ~~...  & \\
 160-140                   &  8128 & 12 & 58.5 & +28 &  4 & 0.25 &  13.7  &
 S            & D:  \hspace{-0.625em} & C                     &  7973 & 1 & ~~...  & ~~...  & \\
 160-147                   &  8134 & 12 & 59.0 & +28 &  8 & 0.30 &  13.7  &
 SABa         & ...                   & ...                   &  5475 & 1 & ~~...  & ~~...  & \\
 160-148A \hspace{-1.08em} &  8135 & 12 & 59.0 & +29 & 35 & 1.12 & (15.0) &
 S pec        & D                     & C                     &  7056 & 1 & ~~S    & ~~VC   & * \\
 160-148B \hspace{-1.08em} &  8135 & 12 & 59.0 & +29 & 35 & 1.12 & (15.0) &
 S pec        & D                     & C                     &  7153 & 1 & ~~...  & ~~...  & * \\
 160-150                   &       & 12 & 59.1 & +28 & 57 & 0.64 &  15.3  &
 S pec        & D:: \hspace{-0.900em} & ...                   &  8909 & 1 & ~~M    & ~~D    & \\
 160-154                   &  8140 & 12 & 59.4 & +29 & 19 & 0.94 &  14.8  &
 Sab          & ...                   & ...                   &  7099 & 1 & ~~M    & ~~D    & \\
 160-159                   &       & 12 & 59.7 & +29 & 31 & 1.11 &  14.9  &
 Sa:          & ...                   & ...                   &  5823 & 1 & ~~...  & ~~...  & \\
 160-160                   &       & 12 & 59.8 & +28 & 29 & 0.47 &  15.5  & 
 pec          & ...                   & [C:] \hspace{-0.5em} &  8311 & 1 & ~~S    & ~~N    & \\
 160-164A \hspace{-1.08em} &       & 13 &  0.2 & +28 & 22 & 0.51 & (16.3) &
 SB:          & ...                   & C                     &  7476 & 2 & ~~...  & ~~...  & * \\
 160-172                   &  8160 & 13 &  0.9 & +28 & 17 & 0.63 &  15.0  &
 S:           & ...                   & (C:) \hspace{-0.5em}  &  6092 & 1 & ~~...  & ~~...  & \\
 160-173                   &  8161 & 13 &  1.0 & +26 & 49 & 1.33 &  15.5  &
 S            & D:  \hspace{-0.625em} & ...                   &  6677 & 1 & ~~...  & ~~...  & \\
 160-176A \hspace{-1.08em} &  8167 & 13 &  1.5 & +28 & 28 & 0.75 & (13.5) &
 Sab          & ...                   & C                     &  7111 & 1 & ~~...  & ~~...  & * \\
 160-178                   &       & 13 &  2.0 & +26 & 56 & 1.35 &  15.3  &
 Sa:          & ...                   & ...                   & 10814 & 1 & ~~...  & ~~...  & \\
 160-179                   &       & 13 &  2.0 & +27 & 34 & 0.99 &  15.5  &
 S: pec       & D:  \hspace{-0.625em} & ...                   &  5523 & 1 & ~~MS   & ~~N    & \\
 160-180                   &       & 13 &  2.0 & +29 &  5 & 1.06 &  15.3  &
 pec          & D:  \hspace{-0.625em} & ...                   &  8050 & 1 & ~~S    & ~~N    & \\
 160-186                   &  8185 & 13 &  3.3 & +28 &  0 & 1.07 &  13.5  &
 Sc           & ...                   & ...                   &  2533 & 1 & ~~MW   & ~~VD   & \\
 160-189A \hspace{-1.08em} &  8194 & 13 &  3.9 & +29 & 20 & 1.45 & (14.0) &
 S            & D:  \hspace{-0.625em} & C                     &  7135 & 2 & ~~...  & ~~...  & * \\
 160-191                   &       & 13 &  4.2 & +29 &  6 & 1.39 &  15.0  &
 pec          & D:  \hspace{-0.625em} & ...                   &  4837 & 1 & ~~S    & ~~N    & * \\
\end{tabular}

 \vspace{\baselineskip} 

 References: \hspace{0.25em}
 1.\ Huchra et al. \shortcite{cfa}\hspace{0.25em} 
 2.\ Nasa Extragalactic Database \hspace{0.25em} 
 3.\ Moss et al. \shortcite{mwi}\hspace{0.25em} 
 4.\ Strauss et al. \shortcite{str}

\end{minipage}    
\end{table*} 

\setcounter{table}{3}

\begin{table*}
\begin{minipage}{165mm}
\caption{\it continued}

Notes on individual objects:

CGCG 522-029A and B: south and north components respectively of double galaxy system.

CGCG 522-086: emission is located $\sim$ 39 arcsec west of a north--south line through the galaxy centre

CGCG 538-043: Emission is double.

CGCG 538-056: Ring galaxy with companion 39 arcsec to east.

CGCG 540-091: possible additional emission $\sim$ 8 arcsec west of a north--south line through the galaxy centre

CGCG 540-112A and B: north and south components respectively of double system.

CGCG 234-079A and B: south and north components respectively of double system. Interacting pair.

CGCG 234-088A: south component of double galaxy system.

CGCG 235-007: emission has two centres.

CGCG 181-006: Ring galaxy with companion 95 arcsec to north-west.

CGCG 159-116: emission has multiple components.

CGCG 160-055: emission is double.

CGCG 160-062A and B: north and south components respectively of double system.

CGCG 160-113A: west component of double galaxy system.

CGCG 160-148A and B: north-east and south-west components respectively of double system. Interacting pair.

CGCG 160-164A: east component of double galaxy system.

CGCG 160-176A: west component of double galaxy system.

CGCG 160-189A: east component of double galaxy system.

CGCG 160-191: emission is possibly double.

 \vspace{\baselineskip}

 Explanations of columns in Table \ref{gsample}

 \indent Column 1.  CGCG number (Zwicky et al. 1960--1968). The
 numbering of CGCG galaxies in field 160 (Abell 1656) which has a
 subfield covering the dense central region of the cluster, follows
 that of the listing of the CGCG in the SIMBAD database. The
 enumeration is in strict order of increasing Right Ascension, with
 galaxies of lower declination preceeding in cases of identical Right
 Ascension.

 \indent Column 2.  UGC number (Nilson 1973)

 \indent Columns 3 and 4.  Right Ascension and Declination (1950.0) of
 the galaxy centre taken from the CGCG.

 \indent Column 5.  Radial distance in Abell radii (Abell 1958) of the
 galaxy with respect to the cluster centre. Positions of the cluster
 centres and values of the Abell radii for the various clusters are
 listed in Table \ref{clusters}.
  
 \indent Column 6. CGCG photographic magnitude. For double galaxies,
 magnitude estimates for individual components obtained by eye from
 PSS are given in parentheses.

 \indent Column 7. Galaxy type taken from UGC or estimated from the PSS

 \indent Column 8. Code indicating that the galaxy appears disturbed, 
  on a 4-rank scale (... [no disturbance], D::, D:, D).

 \indent Column 9. Code indicating that the galaxy has a possible
  nearby companion, on a 4-rank scale (... [no companion], C::, C:,
  C). Square brackets indicate that the companion is likely to be a
  chance superposition, or have negligible tidal interaction with the
  galaxy; parentheses indicate that the probability of the companion
  being a chance superposition, $P$ $>$ 0.05 (see section
  \ref{companion}).

 \indent Columns 10 and 11.  Heliocentric velocity and reference.

 \indent Column 12. A visibility parameter describing how readily the
 H$\alpha$ emission is seen on the plates according to a five-point
 scale (S strong, MS medium-strong, M medium, MW medium-weak, W weak)
 A `?' in this column and column 13 indicates that the galaxy was not
 satisfactorily surveyed for emission for a variety of reasons:
 overlap by an adjacent stellar or galaxy spectrum (CGCG nos. 522-035,
 522-050, 522-071, 522-082, 538-062, 540-115, 540-112B, 540-065,
 540-042); overlap by a ghost image (CGCG no. 181-026); plate defect
 (CGCG no. 160-038); galaxy lies outside the overlap region of the
 plate pair (CGCG nos.  503-030, 503-044, 538-034, 540-036).

 \indent Column 13.  A concentration parameter describing the spatial
 distribution of the emission and contrast with the underlying
 continuum, on a five-point scale (VD very diffuse, D diffuse, N
 normal, C concentrated, VC very concentrated).

 \indent Column 14.  Notes.  An asterisk in this column indicates that
 a note on this galaxy appears below the Table.

 \end{minipage}
 \end{table*}

How do our observed percentage detections of ELGs over the range of
Hubble types using the prism survey, compare with percentages we might
expect to detect based on previous photoelectric and CCD photometry?
To answer this question, we have constructed a comparison sample of
galaxies with photoelectric and CCD measurements of H$\alpha$+[NII]
equivalent width, by combining data given by Kennicutt \& Kent (1983),
Romanishin (1990), and Kennicutt (1992). In order that the sample be
representative of the field, low surface brightness galaxies,
Markarian and Seyfert galaxies, and galaxies from Coma and Abell 1367
were omitted from the sample.  To match our cluster galaxies, the
sample was further restricted in absolute magnitude to $M_{\rm B}^{0}
\le -19$.  In Paper III we established an overall efficiency for the
objective prism technique by comparing its results with photoelectric
observations of galaxies in Abell 1367 and Abell 1656: $\sim 90\%$
complete for $W_{\lambda}
\ge 20$ \AA\/ and $\sim$ $29\%$ complete for $W_{\lambda} \le 20$
\AA. Assuming these detection efficiencies, we derive expected ELG detections
by the prism survey for the comparison sample in the range E to
Sc--Irr.

Note that since we use de Vaucouleurs types for the cluster galaxies,
we also use RC3 (de Vaucouleurs et al. 1991) types for the comparison
sample, following the cautionary statements by Hameed and Devereux
(1999) who note that systematic type differences between the RC3 and
the RSA classification schemes (i.e. those of de Vaucouleurs and
Sandage) may lead to systematic differences in inferred dependence of
star formation on Hubble type.

The dotted histogram in Figure \ref{etdistrib} shows the predicted
fraction of detected galaxies for the comparison sample.  The
detection rate for early type spirals is similar to that predicted
($\chi^{2}$ significance = 0.5 for S0/a--Sb), but there are fewer late
type ELGs detected than expected ($\chi^{2}$ significance =
$6\times10^{-6}$ for Sc and later). It is not immediately clear how to
interpret this. One possibility is that there is reduced star
formation in late type spirals in clusters, but since our later
analysis fails to find this effect (see section \ref{cfcomp}), we
instead suspect that the photographic technique is in fact less
efficient in detecting diffuse emission in late type spirals.  We note
that in Paper II we found a similar result with similar ambiguous
interpretation, while in Paper III there were too few late type
galaxies to attempt a meaningful comparison. Further work is in
progress to better test the detection efficiency of the prism survey.

\begin{figure*}

\epsfxsize=15cm
\hspace*{0.01cm} \epsffile{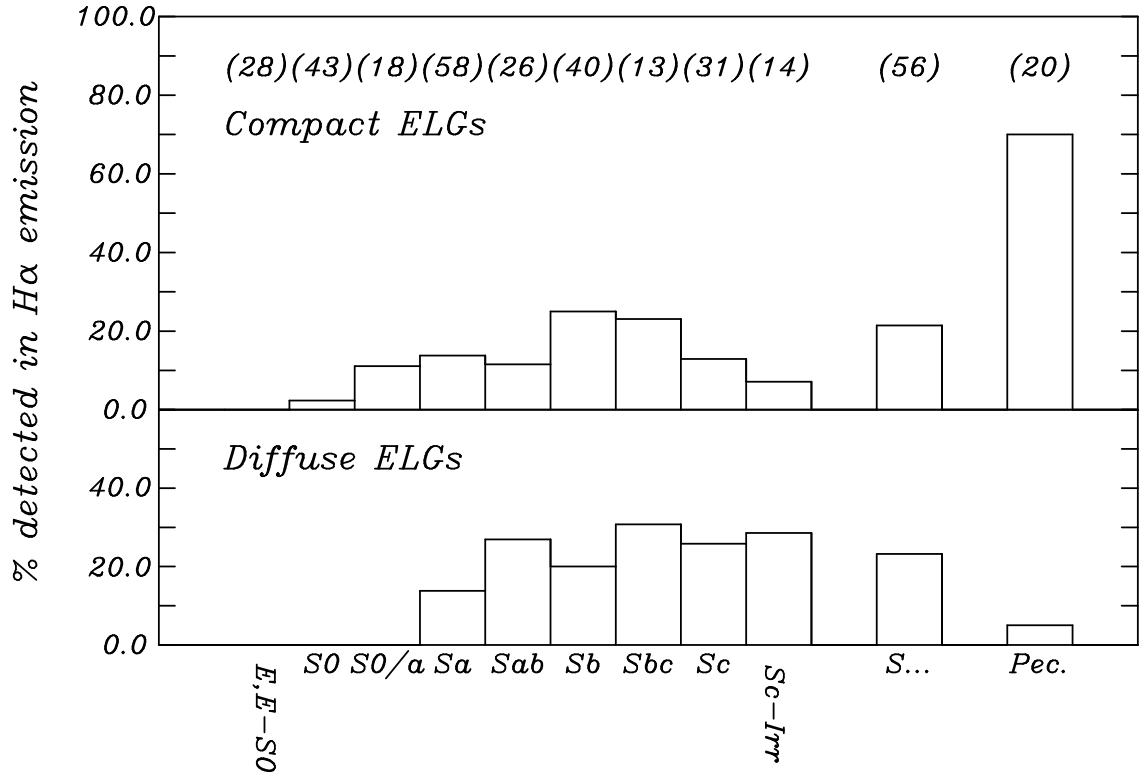}

\caption{\label{etcddistrib}
(Upper panel)
The percentage detection of compact H$\alpha$ + [NII] emission for galaxies
in different Hubble type bins. (Lower panel) As upper panel, for diffuse
emission. The total sample number for each bin
is given in parentheses.
}
\end{figure*}

In Figure \ref{etcddistrib} we compare the Hubble type distribution for
detection of compact ELGs (upper panel) and diffuse ELGs (lower
panel).  Several features of this and the previous Figure may be
noted.  First, the morphological class with the highest
H$\alpha$ detection rate ($\sim 75\%$) are `Peculiar' galaxies, and the
detections are almost exclusively compact emission. In section
\ref{gcomps} below, we suggest that many of these peculiar galaxies
are on-going mergers with associated nuclear starbursts.  Second, the
large but poorly defined class of `S...' galaxies has H$\alpha$
detection rates in each category (all, compact, diffuse) which closely
match those of the `typed' spirals. This supports their inclusion with
the rest of the spirals in the subsequent analysis.  Finally, the small
fraction of early type galaxies (S0,S0/a) detected in H$\alpha$ all
have compact emission, matching the findings of Bennett \& Moss (1998)
for 3 early type galaxies in Abell 1060. This suggests a tidal or
post-merger origin for the emission in these systems (see section
\ref{cenviron}).

\subsection {Bars}
\label{bars}

Does the presence of a bar influence star formation?  For the 128
galaxies with bar information, Kendall rank tests yield $0.3\sigma$,
$3.6\sigma$ and $-3.2\sigma$ for correlation strengths between the
presence of a bar and the emission line categories ELG(all),
ELG(compact), ELG(diffuse) respectively.  Thus, it seems that barred
galaxies are more likely to have compact emission than unbarred
spirals, but less likely to have diffuse emission.  While spiral stage
is not expected to influence these correlations (because there is no
significant dependence of bars on spiral stage), galaxy disturbance
may influence them, since compact emission and disturbance are
strongly linked (see below).  Accordingly we have calculated the
corresponding partial rank correlation coefficients for the case in
which galaxy disturbance is partialled out, and estimated their
significance levels using a bootstrap resampling technique provided by
Biviano (private communication).  The resulting significance levels
are 0.7$\sigma$, 3.1$\sigma$ and -3.0$\sigma$ for correlation
strengths between the presence of a bar and the emission line
categories ELG(all), ELG(compact), ELG(diffuse) respectively. These
results confirm, in agreement with Paper III, that while there is no
overall enhancement of emission in barred spirals, they tend to have
compact emission.  The same results indicate that unbarred spirals
tend to have diffuse emission, although this may at least partly be
due to the presence of diffuse emission on the prism plates being
overlooked for a number of galaxies due to the dominance of compact
emission.

\subsection {Disturbance}
\label{disturb}

For the full cluster sample, a Kendall rank test between disturbance
and the emission line categories ELG(all), ELG(compact), ELG(diffuse)
give correlation significances of $5.4\sigma$, $8.7\sigma$ and
$-2.2\sigma$ respectively. As noted above (section \ref{htype}), most
galaxies classified as peculiar also have compact emission.  Omitting
these types leaves a sample of 270 spirals for which a Kendall rank
test shows similar correlations: $4.4\sigma$,  $7.5\sigma$ and
$-1.6\sigma$ respectively.  Thus, a disturbed galaxy morphology is a
strong predictor of compact emission.  This correlation is the more
striking since indications of a disturbed morphology are generally
taken from the {\it outer} features of the galaxy.  The obvious
explanation for this correlation is tidally-induced star formation
which is discussed further in the section \ref{gpdiscuss} below. These
results for the full sample echo our previous work in Papers II and III
with more limited samples.

\subsection {Galaxy companions and mergers}
\label{gcomps}

Although a Kendall rank test for the companion parameter with emission
yields no significant result for the combined sample of all ELGs
(significance of -0.3$\sigma$), there are significant correlations in
opposite senses for compact and diffuse emission taken separately
(significances of 3.1$\sigma$ and -3.5$\sigma$ respectively). The
exclusive tendency for compact ELGs to have tidal companions, further
confirms an explanation of tidally-induced star formation for much of
this emission, some of which is caused by local galaxy--galaxy
interactions.  

On the other hand, as noted above (section \ref{htype}) a very high
percentage ($\sim$ 70\%) of galaxies classified as peculiar have
compact emission.  However these galaxies show no tendency to have
tidal companions (Kendall rank test significance = 1.4$\sigma$).  A
natural explanation of these results is that the peculiars are
predominantly on-going mergers, in which the companion is already
indistinguishable from its merger partner.  These then represent a
later stage of close double, interacting systems, many examples of
which are found in the clusters with tidally-induced star formation
(e.g. CGCG nos. 540-112, 234-079, 127-025, 97-079, 97-092, 127-051,
160-113 and 160-148).  For the remaining sample of spiral galaxies
alone, a Kendall rank test shows an even stronger correlation between
the companion parameter and compact emission (significance of
4.6$\sigma$) consistent with this picture.

Interestingly, there is an {\it anticorrelation} between diffuse
H$\alpha$ emission and a tidal companion, although, as for the observed
anticorrelation of diffuse emission and the presence of a bar (see
section \ref{bars}), this may at least partly be due to the presence of
diffuse emission being overlooked in cases where compact emission is
dominant.

\subsection {Starburst and disk emission}
\label{gpdiscuss}

Kennicutt (1998) notes that large-scale star formation takes place in
two very distinct physical environments, viz. in the extended disks of
spiral and irregular galaxies, and in compact, dense gas disks in the
centres of galaxies.  Line emission associated with star formation in
the two regions have very different dependencies on galaxy morphology.
In particular, circumnuclear emission has a strong dependence on a
barred structure but weak dependence on Hubble type, while the opposite
is true for disk emission.  In addition, a clear causal link between
strong nuclear starbursts and tidal interactions and mergers has been
established by numerous observations (e.g. Keel et al. 1985; Bushouse
1987; Kennicutt et al. 1987; Wright et al. 1988) consistent with
predictions of numerical simulations (e.g. Noguchi 1988; 
Hernquist 1989; Mihos \& Hernquist 1996).  For
nearby samples of interacting galaxies, the H$\alpha$ emission is
typically 3--4 times stronger than for isolated spirals.

In view of the above, the obvious and most compelling interpretation
of the distinction between compact and diffuse emission in our survey
sample is that of circumnuclear starburst and disk emission
respectively.  As has been seen, compact emission is generally centred
on the nucleus of the galaxy and is of smaller spatial extent (median
diameter $\sim$ 2.5 kpc, Paper II), and correlates with a barred
structure, all of which is typical of circumnuclear emission.
Furthermore compact emission is strongly correlated with a disturbed
morphology and with the presence of a nearby companion, strongly
suggesting that much of this emission is indeed due to tidally-induced
nuclear starbursts.  
Finally there is no significant dependence of compact emission on 
Hubble stage from Sa--Sc.
By contrast, diffuse emission has a greater
spatial extent closer to that expected for disk emission (median
diameter $\sim$ 7 kpc, Paper II) and is not correlated either with a
barred structure or a disturbed morphology. There is no apparent 
dependence of emission on Hubble stage, but this is not considered
significant because of uncertainties associated with detection of
diffuse emission (cf. section \ref{htype}).

Despite the above considerations, could the compact emission be due to
non-stellar emission?  Ho, Filippenko \& Sargent (1997a) have
extended earlier studies to show that AGN and LINER (Heckman 1980)
emission is very common, particularly in early-type spirals of
which 60\% show this non-stellar activity.  However galaxies in their
sample have low H$\alpha$ luminosity (median $L_{H\alpha} \sim
10^{39}$ erg s$^{-1}$). By contrast the H$\alpha$ luminosities of the
ELGs in our survey sample are higher ($10^{40} \la L_{H\alpha} \la
5 \times 10^{41}$ erg s$^{-1}$), more typical of lower luminosity starburst
emission (Balzano 1983). Moreover Ho, Filippenko \& Sargent (1997b)
show that whereas bars enhance nuclear star formation in their sample,
there is no corresponding enhancement of AGN activity.  Finally,
despite the fact that a few of our ELGs are classed as Seyferts
(viz. CGCG nos. 126-110, 522-081, 540-064, 540-103 and 160-148A) a
spectral survey of ELGs in Abell 1367 (Moss \& Whittle, unpublished)
has confirmed that emission for most of these galaxies resemble HII
regions and not AGN or LINERs.  For these reasons, it is considered
unlikely that most compact emission has a non-stellar origin.  In what
follows, we assume that both compact and diffuse emission are due to
photoionisation by massive young stars, and investigate how the
corresponding circumnuclear starburst and disk star formation varies
within a cluster and field environment.

\section {H$\alpha$ detection and cluster environment}
\label{cenviron}

\subsection {Introduction}
\label{ceintro}

Using the full sample of surveyed galaxies, we compare emission
detection rates in cluster and field spirals in a manner which
overcomes three principal limitations of earlier studies (Papers II and
III).  First, we observe our own field sample in an identical manner to
the cluster sample, unlike our earlier work which used a field sample
observed by photoelectric photometry, possibly introducing systematic
biases.  Second, our field sample is greatly enlarged from that used
for Paper III.  Third, whereas in Paper III we made the cluster/field
comparison for only a single cluster, here we make use of the entire
survey sample which includes a full range of cluster types.

Before proceeding to the comparison, it is first necessary to
consider how this may be affected by the use of Hubble types
and by field galaxy contamination.

\subsubsection {Hubble type and bulge/disk ratio in cluster/field comparisons}
\label{cfcspirals}

We have chosen to use Hubble types to normalise galaxy samples when
comparing cluster and field galaxies.  Several authors have noted that
the Hubble classification system was based mainly on nearby field
spirals and may not be adequate to describe environmentally altered
galaxies in dense environments (e.g. van den Bergh, 1976, 1997;
Koopmann \& Kenney 1998). In particular, one characteristic used to
determine Hubble type is the resolution of the spiral arms, which is
itself related to star formation. Thus, for example, a decrease in
the disk star formation which also shifts a galaxy to an earlier
Hubble type may not be detected in any comparison of field and cluster
spirals (Hashimoto et al. 1998).

Some authors (e.g.  Hashimoto et al. 1998, Balogh et al. 1998) have
instead adopted a measure of bulge to disk ratio (B/D) as a less
subjective and star formation-contaminated normalisation parameter.
However, the use of B/D ratio may introduce other problems. First, the
relation between B/D and Hubble T-type has sufficient scatter (Baugh,
Cole \& Frenk 1996; Simien \& de Vaucouleurs 1986; de Jong 1995) that a galaxy
with B/D = 1, for example, could lie anywhere in the range S0--Sbc.  It
is still not known whether the scatter in the relation is observational
or real (Baugh et al. 1996).  Second, because of this scatter in the
B/D vs T-type relation, S0 galaxies will be included in both field and
cluster samples, and so an increase in the S0/S ratio in clusters can
lead to a perceived reduction in star formation rate.  Thirdly, the B/D
ratio itself may depend on star formation.  A change in star formation
rate in either the disk and/or the circumnuclear regions will change the
B/D ratio (Balogh et al. 1998).

Thus, although there may be no better alternative to using the Hubble
type for cluster/field comparison, the use of Hubble types introduces a
possible limitation to our study, at least for {\it disk} emission.  In
what follows, we find no difference between disk emission in field and
cluster samples. It remains unclear to what extent this result may
represent a limitation of the method, rather than a true comparison of
the two samples.  By contrast, we do expect a comparison based on
Hubble type to be sensitive to differences of {\it circumnuclear} star
formation.  Such star formation has little or no relation to Hubble
type (Kennicutt 1998), and, consistent with this, is not expected to
affect the type classification.

\subsubsection {Sample contamination by field galaxies}
\label{scfgals}

Any comparison of cluster and field galaxies needs to allow for
possible contamination of the cluster sample by projected field
galaxies.  This contamination is more severe for late-type galaxies
(such as our own sample) than early-type galaxies because late type
galaxies are less common in clusters than in the field.

We have attempted to estimate the contamination effect on our cluster
spiral sample for a series of concentric zones for all 8 clusters (see
below).  The estimates are useful in several ways. First, they
confirm that contamination by field projection is not
important at least in the cluster centres and for regions of high
galaxy surface density. Second, the estimates can be used to select
zones for a true comparison of cluster and field samples.  And finally,
the space density of the central zone of each cluster
can be used to rank the clusters for tests of the
dependence of emission and other galaxy properties on cluster type.

We estimate field contamination as follows. First, for each cluster
(except Abell 569 which is double), we consider four concentric annular
zones, 1--4: $0.0-0.5{\rm r}_{\rm A}$; 
$0.5{\rm r}_{\rm A}-1.0{\rm r}_{\rm A}$;
$1.0{\rm r}_{\rm A}-1.5{\rm r}_{\rm A}$; 
and $1.5{\rm r}_{\rm A}-3.0{\rm r}_{\rm A}$, 
where ${\rm r}_{\rm A}$ is the
Abell radius.  We assume that all
galaxies in the outermost annulus, 
$1.5{\rm r}_{\rm A} - 3.0{\rm r}_{\rm A}$, are field
galaxies.  For each cluster, Table \ref{scontam} gives the total number
of spirals, $n_{s}$, in zones 1--3, and the total number of galaxies of
{\it all} types, $n_{t}$, in the outermost zone.  For Abell 569 the
principal component is situated at the cluster centre, and a secondary
component lies approximately 1\fdg5 north.  Values of $n_{s}$ are given
only for regions of radius $0.5{\rm r}_{\rm A}$ centred on each of the two
subclusters.

\begin{table*}
\centering
\caption{\label{scontam} Field galaxy contamination of cluster spiral sample}

\begin{tabular}{lrrrcrrrcrrrcc} \hline
Cluster & \multicolumn{3}{c}{Zone 1} && \multicolumn{3}{c}{Zone 2} && 
\multicolumn{3}{c}{Zone 3} && \multicolumn{1}{c}{Zone 4} \\ 
 & \multicolumn{3}{c}{(0.0--0.5${\rm r}_{\rm A}$)} && 
   \multicolumn{3}{c}{(0.5${\rm r}_{\rm A}$--1.0${\rm r}_{\rm A}$)} &&
   \multicolumn{3}{c}{(1.0${\rm r}_{\rm A}$--1.5${\rm r}_{\rm A}$)} &&
   \multicolumn{1}{c}{(1.5${\rm r}_{\rm A}$--3.0${\rm r}_{\rm A}$)} \\ 
 \cline{2-4} \cline{6-8}
 \cline{10-12} \cline{14-14}
 & \multicolumn{1}{c}{$n_{s}$} & \multicolumn{1}{c}{$n_{fs}$} &
   \multicolumn{1}{c}{$p_{fs}$} &~&  
   \multicolumn{1}{c}{$n_{s}$} & \multicolumn{1}{c}{$n_{fs}$} &
   \multicolumn{1}{c}{$p_{fs}$} &~&  
   \multicolumn{1}{c}{$n_{s}$} & \multicolumn{1}{c}{$n_{fs}$} &
   \multicolumn{1}{c}{$p_{fs}$} &~&  
   \multicolumn{1}{c}{$n_{t}$} \\ \hline \hline
 Abell 262 &  16 &  4 &  22\% && 25 & 11 &  42\% && 16 & 18 & 
  [100\%] \hspace*{-0.6em} && \hspace{-0.85em} 155 \\
 Abell 347 &  15 &  3 &  18\% && 12 &  8 &  67\% && 13 & 13 &  
  [100\%] \hspace*{-0.6em} && \hspace{-0.85em} 119 \\
 Abell 400 &   6 &  1 &  15\% &&  6 &  3 &  45\% &&  2 &  5 &  
  [100\%] \hspace*{-0.6em} &&  40 \\
 Abell 426 &  14 &  2 &  13\% &&  9 &  5 &  61\% && 17 &  9 &   54\%  &&  81 \\
 Abell 569 &  14 &  2 &  12\% &&    &    &       &&    &    &         &&  77 \\
 Abell 569N&   6 &  2 &  29\% &&    &    &       &&    &    &         &&  77 \\
 Abell 779 &   6 &  1 &  21\% &&  7 &  4 &  54\% &&  9 &  6 &   70\%  &&  56 \\
 Abell 1367&  19 &  3 &  17\% &&  9 & 10 &   [100\%] \hspace{-0.6em} && 
 20 & 16 &   80\%  && \hspace{-0.85em} 141 \\
 Abell 1656&  11 &  2 &  15\% && 14 &  5 &  34\% && 12 &  8 &   67\%  &&  71 \\ \hline
\end{tabular}
\end{table*}

First we need an estimate of the spiral fraction in zone 4, since most
CGCG galaxies in this zone have not been typed.  As shown below (see Table \ref{spden}), the
true space density (i.e. after field correction) in zone 3 for each of
the 4 least rich clusters (Abell 262, 347, 400 and 779), is essentially zero.
Thus galaxies in these zones can be considered projected
(supercluster) field galaxies. From these zones, and a total of 78
typed CGCG galaxies in zone 4 of Abell 1367 (cf. Paper III), we
measure a spiral fraction of 61\%.  This value was adopted for the
spiral fraction for zone 4 of all the clusters, and used to estimate
the number of projected field spirals, $n_{fs}$, and the percentage
contamination, $p_{fs}$, for zones 1--3 in each of the clusters (see
Table \ref{scontam}).

Note that although there is considerable contamination for zone 3
(outside the nominal limit of the clusters), and significant
contamination for zone 2 for most of the clusters ($p_{fs} \ga 50\%$),
in zone 1 the contamination is generally low ($p_{fs} \sim 17\%$). This
gives us confidence that a valid comparison is possible between cluster
and field spirals in the survey sample, at least for this central
zone.

Next, following a procedure similar to that of Wallenquist (1960) and
assuming spherical symmetry for the cluster and uniform density within
each annulus, the apparent space densities, $d_1$, $d_2$, and $d_3$ in
each zone in units of galaxies ${\rm r}_{\rm A}^{-3}$, are given by:

\[d_{3}=\frac{n_{3}^{c}}{\frac{4}{3} \pi ({r_{3}}^{2}-{r_{2}}^{2})^{\frac{3}{2}}} \]
\[d_{2}=\frac{n_{2}^{c}-Xd_{3}}{\frac{4}{3} \pi 
({r_{2}}^{2}-{r_{1}}^{2})^{\frac{3}{2}}} \]
\[d_{1}=\frac{n_{1}^{c}-Yd_{3}-Zd_{2}}{\frac{4}{3} \pi {r_{1}}^{3}} \]
where
\[X=\frac{4}{3} \pi ({r_{3}}^2-{r_{1}}^2)^{\frac{3}{2}} 
-\frac{4}{3} \pi ({r_{3}}^2-{r_{2}}^2)^{\frac{3}{2}} 
-\frac{4}{3} \pi ({r_{2}}^2-{r_{1}}^2)^{\frac{3}{2}} \]
\[Y=\frac{4}{3} \pi {r_{3}}^{3}
-\frac{4}{3} \pi ({r_{3}}^2-{r_{1}}^2)^{\frac{3}{2}} 
-\frac{4}{3} \pi {r_{2}}^{3}
+\frac{4}{3} \pi ({r_{2}}^2-{r_{1}}^2)^{\frac{3}{2}} \]
\[Z= \frac{4}{3} \pi {r_{2}}^{3}
-\frac{4}{3} \pi {r_{1}}^{3}
-\frac{4}{3} \pi ({r_{2}}^2-{r_{1}}^2)^{\frac{3}{2}} \]

and $n_{1}^{c}$, $n_{2}^{c}$ and $n_{3}^{c}$ are the total
numbers of CGCG galaxies of all types in zones 1, 2 and 3 respectively,
corrected for projected field galaxies.  Taking $r_1=0.5$, $r_2=1.0$,
$r_3=1.5$, we have $X=3.2730$, $Y=0.8214$, $Z=0.9445$, which yield the
apparent space densities given in Table \ref{spden}.  For the double
cluster Abell 569 we have adopted a simplified procedure.  The galaxy
count for zone 1 for the two cluster components was corrected only for
projected field galaxies, and the resulting corrected count was used to
determine values of $d_{1}$ given in the Table.

\begin{table*}
\centering
\caption{\label{spden} Cluster zonal space densities}
\begin{tabular}{lrrrcrrrcrrr} \hline
Cluster& \multicolumn{1}{c}{$n_{1}$} & 
         \multicolumn{1}{c}{$n_{2}$} & 
         \multicolumn{1}{c}{$n_{3}$} && 
         \multicolumn{1}{c}{$d_{1}$} & 
         \multicolumn{1}{c}{$d_{2}$} & 
         \multicolumn{1}{c}{$d_{3}$} && 
         \multicolumn{1}{c}{$\rho_{1}$} & 
         \multicolumn{1}{c}{$\rho_{2}$} & 
         \multicolumn{1}{c}{$\rho_{3}$} \\ \cline{6-8} \cline{10-12}
&&&&& \multicolumn{3}{c}{\raisebox{-0.5ex}{(${{\rm r}_{\rm A}}^{-3}$)}} &&
      \multicolumn{3}{c}{\raisebox{-0.5ex}{(${\rm Mpc}^{-3}$)}} \\ 
 \hline \hline
 Abell 262 & 44 & 45 & 23 &&  54.7 & 10.2 & 0.0 && 2.10 & 0.39 & 0.00 \\
 Abell 347 & 23 & 24 & 21 &&  28.4 &  4.0 & 0.0 && 1.62 & 0.23 & 0.00 \\
 Abell 400 & 16 &  9 &  3 &&  24.7 &  1.7 & 0.0 && 3.09 & 0.21 & 0.00 \\
 Abell 426 & 41 & 27 & 36 &&  62.8 &  2.3 & 3.6 && 3.76 & 0.14 & 0.21 \\
 Abell 569 & 24 &    &    &&  40.4 &      &     && 2.30 &      &       \\
 Abell 569N& 24 &    &    &&  40.4 &      &     && 2.30 &      &       \\
 Abell 779 & 19 & 15 &  9 &&  26.5 &  3.1 & 0.0 && 1.68 & 0.20 & 0.00 \\
 Abell 1367& 67 & 21 & 35 && 115.4 &  0.1 & 1.5 && 7.39 & 0.01 & 0.10 \\
 Abell 1656& 85 & 62 & 37 && 122.6 & 15.7 & 4.1 &&10.72 & 1.30 & 0.35 \\ \hline
\end{tabular}
\end{table*}

The apparent space densities are not directly comparable because the
magnitude limit of the CGCG catalogue, $m_{\rm p}=15.7$, corresponds to
different absolute magnitude limits, $M_{\rm B}$, depending on cluster
distance modulus and Galactic reddening.  Using the conversion of
$m_{\rm p}$ to absolute magnitude given in section \ref{mags}, and adopting
a common limit, $M_{\rm B} \la -19.5$, we obtain the true space densities,
$\rho_1$, $\rho_2$, $\rho_3$ (in units, galaxies ${\rm Mpc}^{-3}$,
see Table \ref{spden}). These space densities will be used for
the ranking of clusters in section \ref{ctype} below.

\subsection {Cluster/field parameters}
\label{cfparam}

We have used three parameters to compare the incidence of star
formation in clusters and field spirals: projected radial distance from the
cluster centre, $R$; local galaxy surface density, $\Sigma$; and
cluster type, $CT$, determined by the central galaxy density.  These
parameters are, of course, closely related:  $\Sigma$ and $CT$ are
strongly correlated, while $R$ and $\Sigma$ are strongly
anticorrelated. Before using these parameters (see section
\ref{cfcomp}), we briefly define them.

\subsubsection {Projected radial distance from cluster centre, $R$}
\label{rden}

Using the projected radial distance, $R$, for each surveyed galaxy,
measured in Abell radii, galaxies in each of the surveyed clusters
(except the double cluster Abell 569) were stacked into a single
`synthetic' cluster.  For the purpose of Kendall rank tests, the
survey sample was divided into ten radial bins, each with
approximately equal populations ($n \sim 32$).

Use of the radial distance parameter has obvious limitations. The
method neglects azimuthal variations in galaxy density, as well
as systematic variations in cluster properties.  Nevertheless any
systematic change in emission properties of spirals from the field to
a cluster environment might be expected to show a systematic change with
$R$.

\subsubsection {Local galaxy surface density, $\Sigma$}
\label{sdden}

To define a local galaxy surface density parameter, $\Sigma$, we
follow the procedure used by Dressler (1980).  First, for each
surveyed galaxy, the 10 nearest projected CGCG neighbours are
identified and the distance to the tenth nearest defines the
radius of a circle. After correction for field galaxy contamination,
the galaxy surface density in this circle is calculated.  If the
estimated number of projected field spirals in the circle is $\ge 10$,
the surface density is set to zero.  A correction is made for the
different absolute magnitude limits of the galaxy counts for each
cluster.  The final value of the local surface density, $\Sigma$, is
the number of galaxies, $M_{\rm B} \le -19.5$, Mpc$^{-2}$.  For the
purpose of Kendall rank tests, surveyed galaxies were divided into
discrete bins covering the range of $\Sigma$.  Galaxies with values of
$\Sigma = 0$ were gathered in one bin ($n=132$) and remaining galaxies
were grouped in 9 bins according to surface density with approximately
equal populations ($n \sim 21$).
 
\subsubsection {Cluster type, $CT$}
\label{ctype}

We have ranked each cluster according to its central galaxy space
density, $\rho_{1}$, defined as the mean space density of galaxies, $M_{\rm B}
\le -19.5$, Mpc$^{-3}$ within the central region $r \le 0.5{\rm
r}_{\rm A}$ (see Table \ref{spden}). In addition we assign the lowest
rank to field (supercluster) spirals which comprise surveyed galaxies
with $> 1.5 {\rm r}_{\rm A}$ together with those in zone 3 of Abell
262, 347, 400 and 779 (see section \ref{scfgals} above).  Cluster
galaxies were taken as those surveyed galaxies with $r \le 1.0 {\rm
r}_{\rm A}$.  Ranks were assigned as follows: rank 1, field spirals as
above; rank 2 ($\rho_{1} \sim 2$ Mpc$^{-3}$), Abell 262, 347, 569,
779; rank 3 ($\rho_{1} \sim 3$ Mpc$^{-3}$), Abell 400; rank 4
($\rho_{1} \sim 4$ Mpc$^{-3}$), Abell 426; rank 5 ($\rho_{1} \sim 7$
Mpc$^{-3}$), Abell 1367; and rank 6 ($\rho_{1} \sim 11$ Mpc$^{-3}$),
Abell 1656.

\subsection {Comparison of emission detection rates for different
environments}
\label{cfcomp}

In considering the dependence of emission detection rates on different
environments using the parameters $R$, $\Sigma$ and $CT$ defined
above, we need to ensure that any significant correlations which arise
are not spuriously due to indirect dependencies on other variables.
To assess this, we first consider Kendall rank tests between these 3
parameters and several possibly relevant galaxy properties, viz.
Hubble type, bar, disturbed morphology, and the incidence of galaxies
classified as peculiar. For these tests (and subsequent tests of
emission detection rates on environment), the sample was restricted to
galaxies whose known radial velocity was not greater than 3$\sigma$
from the cluster mean.  Results of these tests are given in Table
\ref{kprops}.  For this, and Tables \ref{kem} and \ref{kspem} below,
test results are given as the significance in units of $\sigma$ with
the sample number in parentheses.

First, it is seen that, as noted above, there is no significant
correlation between Hubble stage and either $R$ or $\Sigma$, and a
possible weak anticorrelation with $CT$.  Thus in considering systematic
correlations of emission detection rate with either $R$ or $\Sigma$,
the effect of Hubble stage can be neglected.  For the parameter $CT$,
the effect of the systematic variation of Hubble stage is to decrease the
likelihood of emission for cluster as compared to field galaxies.
However, in what follows, we are concerned with an {\it increase} of
emission detection for cluster galaxies, and the effect of Hubble
stage is thus to make this increase even more significant.

Second, it is seen that there is no significant correlation of a
barred structure with either $R$, $\Sigma$ or $CT$.  In section
\ref{bars} above, it was noted that there is a correlation of
compact emission with a barred structure.  Below, we will note a
strong enhancement of compact emission for cluster as compared to
field spirals.  The lack of correlation between barred structure and a
cluster environment shows that the enhanced compact emission cannot be
due to an increase in barred structure in cluster spirals.  

Next, the results show a possible weak correlation of a disturbed
morphology with local galaxy surface density.  This effect can most
simply be attributed to enhanced tidal effects on spirals in higher
density regions.

Finally, it is seen that peculiar galaxies are more likely to be found in 
higher surface density regions and in richer clusters.  Since a large
percentage of these galaxies ($\sim$ 70\%) show compact emission, a
corresponding increase of compact emission in the cluster sample as 
compared to the field is to be expected.

In Table \ref{kem} we give Kendall rank test results for emission in
the three categories, ELGs(all), ELGs(compact) and ELGs(diffuse) with
each of the parameters $R$, $\Sigma$ and $CT$.  Since for these
results and subsequent results given in Table \ref{kspem}, the most
significant correlations are found for $\Sigma$ and $CT$, we will
discuss these.  Results for the parameter $R$ are generally indicative
of similar effects found by the other two parameters, but are much
weaker and accordingly of less interest.

From Table \ref{kem}, it is seen that there is a significant
correlation of emission detection rate with $CT$ (significance level,
3.2$\sigma$) and some suggestion of such a correlation with $\Sigma$
(significance level, 2.6$\sigma$).  Galaxies of types Sa and later
thus are more likely to have emission in clusters with higher than
lower central density.  This result is a surprising one, and the
opposite of that expected on the basis of cluster spiral gas content,
and its significance is perhaps even greater than the test results
indicate, due to a weak correlation of Hubble stage with cluster
density which contributes to lowering the emission detection rate for
galaxies in the most dense clusters.

Furthermore the correlation of emission detection rate with $\Sigma$
and $CT$ is seen to be entirely due to a very significant correlation
of compact emission with these parameters (3.9$\sigma$ and 5.3$\sigma$
respectively).  This enhancement of compact emission in cluster
galaxies as compared to the field is not due simply to an increased
likelihood of `peculiar' galaxies being found in the cluster
environment.  In Table \ref{kem} we show Kendall rank test results
between ELGs(all), ELGs(compact) and ELGs(diffuse) and $R$, $\Sigma$
and $CT$ for the spiral sample alone (excluding galaxies classed as
`peculiar').  There is a similar correlation of emission detection
rate with both $\Sigma$ and $CT$, as for the full sample, and the
increase in emission in regions of higher density and for clusters of
higher central density is again seen to be entirely due to enhanced
compact emission.  As noted above, the enhanced compact emission in
cluster spirals cannot be due to an increase in a barred structure for
these galaxies.  Rather, we conclude that it is due to low luminosity
circumnuclear starbursts due to increased tidal interactions in the
cluster environment.  This view is supported by the fact that a
Kendall rank test shows that both the most disturbed spirals and peculiar
galaxies are preferentially found in clusters of higher central
density (significance levels of 3.2$\sigma$ and 3.8$\sigma$
respectively).

The enhancement of compact emission in cluster as compared to field
spirals has been shown from the correlation of the emission detection
rate with both local galaxy surface density and with cluster type,
ranked according to central galaxy density.  Is the emission
enhancement entirely due to local galaxy surface density, with the
observed correlation with cluster type simply due to a greater
proportion of galaxies in the more dense clusters being situated in
regions of higher surface density?  Or is there a `cluster effect',
such that galaxies in a region of a given surface density in more
dense clusters, are more likely to have compact emission than galaxies
in a region of the same surface density in less dense clusters?  A
Kendall partial rank test of the correlation of compact emission with
cluster type for the case in which local galaxy surface density is
partialled out yields a significance level = 3.3$\sigma$.  It thus
appears that there is indeed a `cluster effect' and that galaxies in a
region of a given surface density in a more dense cluster are more
likely to have compact emission than galaxies in a region of the same
surface density in a less dense cluster.  The implications of this
result will be discussed further in section \ref{discuss} below.

\begin{table}
\centering
\caption{\label{kprops} Kendall rank tests: 
cluster/field and galaxy properties}

\begin{tabular}{l@{\hspace{0.5em}}r@{\hspace{1.0em}}r@{\hspace{0.5em}}r
@{\hspace{1.0em}}r@{\hspace{0.5em}}
rr@{\hspace{0.5em}}r@{\hspace{1.0em}}r} \hline
 & \multicolumn{6}{c}{Spirals only} & \multicolumn{2}{c}{Peculiar} \\ 
\cline {2-7}
 & \multicolumn{2}{c}{\raisebox{-0.5ex}{Hubble}} & 
 \multicolumn{2}{c}{\raisebox{-0.5ex}{Bar}} & 
 \multicolumn{2}{c}{\raisebox{-0.5ex}{Disturbed}} & \multicolumn{2}{c}{type} \\
 & \multicolumn{2}{c}{stage} &&& \multicolumn{2}{c}{morphology} & \\ 
\hline \hline
R & 0.6$\sigma$ \hspace{-0.95em} & (148) & 
    1.6$\sigma$ \hspace{-0.95em} & (107) &
    $-1.4\sigma$ \hspace{-0.95em} \hspace{-0.75em} & (235) \hspace{0.5em} &
  $-1.3\sigma$ \hspace{-0.95em} & (257) \\     
$\Sigma$ & $-0.5$ & (148) & $-0.3$ & (107) & 2.2 \hspace{-0.75em}
& (235) \hspace{0.5em} & 1.9 & (257) \\
CT & $-2.2$ & (124) & $-1.4$ & (86) & 1.5 \hspace{-0.75em}
& (194) \hspace{0.5em} & 3.8 & (210) \\ \hline
\end{tabular}
\end{table}

In Table \ref{kspem} we give results of Kendall rank tests between
ELGs(all), ELGs(compact) and ELGs(diffuse) and $R$, $\Sigma$ and $CT$
for the spiral sub-groups, Sa and Sab, Sb and Sbc, and Sc--Irr.  It is
seen that for each of the sub-groups, there is the same increase of
compact emission with higher surface density regions and more dense
clusters for as for the full sample of spirals combined. Of particular
interest is the very significant correlation of emission detection
rate with increasing cluster density for Sc--Irr galaxies.  In section
\ref{htype}, it was seen that surveyed spirals of these types have a
lower detection rate than expected from photoelectric and CCD
photometry.  We can conclude that it is unlikely that this lower
detection rate is due to any lessened emission from cluster as
compared to field spirals.  Rather it is more likely, as previously 
suggested (section \ref{htype}), 
that this lower detection rate is due to non-detection of
diffuse disk emission in the low surface brightness disks of these
galaxies by the photographic survey.  Work is in progress
to verify this conclusion.

\begin{table*}
\centering
\caption{\label{kem} Kendall rank tests: 
cluster/field and emission detection}

\begin{tabular}{lrrrrrrcrrrrrr} \hline
 & \multicolumn{6}{c}{All sample} &~& \multicolumn{6}{c}{Spirals only} \\ 
\cline {2-7} \cline{9-14}
 & \multicolumn{2}{c}{\raisebox{-0.5ex}{ELGs}} & 
 \multicolumn{2}{c}{\raisebox{-0.5ex}{ELGs}} & 
\multicolumn{2}{c}{\raisebox{-0.5ex}{ELGs}} &&
\multicolumn{2}{c}{\raisebox{-0.5ex}{ELGs}} & 
\multicolumn{2}{c}{\raisebox{-0.5ex}{ELGs}} & 
\multicolumn{2}{c}{\raisebox{-0.5ex}{ELGs}} \\
 & \multicolumn{2}{c}{(all)} & \multicolumn{2}{c}{(compact)} &
\multicolumn{2}{c}{(diffuse)} &&
\multicolumn{2}{c}{(all)} & \multicolumn{2}{c}{(compact)} &
\multicolumn{2}{c}{(diffuse)} \\
\hline \hline
R &$-1.0\sigma$ \hspace{-0.95em} & (257) & 
   $-2.6\sigma$ \hspace{-0.95em} & (257) &
    1.4$\sigma$ \hspace{-0.95em} & (257) &&
   $-2.4\sigma$ \hspace{-0.95em} & (237) &
   $-1.9\sigma$ \hspace{-0.95em} & (237) &     
   $-1.1\sigma$ \hspace{-0.95em} & (237) \\     
$\Sigma$ & 2.6 & (257) & 3.9 & (257) & $-0.7$ & (257) &&
           1.8 & (237) & 2.5 & (237) & $-0.2$ & (237) \\
CT       & 3.2 & (210) & 5.3 & (210) & $-1.3$ & (210) &&
           2.3 & (195) & 4.0 & (195) & $-0.8$ & (195) \\ \hline      
\end{tabular}
\end{table*}

\begin{table*}
\centering
\caption{\label{kspem} Kendall rank tests: 
cluster/field spiral sub-types and emission detection}

\begin{tabular}{l@{\extracolsep{0.7em}}r@{}r@{}r@{}r@{}
r@{}r@{}c@{}r@{}r@{}r@{}r@{}r@{}r@{}c@{}r@{}r@{}r@{}r@{}r@{}r} \hline
 & \multicolumn{6}{c}{Sa,Sab} &~& \multicolumn{6}{c}{Sb,Sbc} &~&
\multicolumn{6}{c}{Sc--Irr} \\
\cline {2-7} \cline{9-14} \cline{16-21}
 & \multicolumn{2}{c}{\raisebox{-0.5ex}{ELGs}} & 
   \multicolumn{2}{c}{\raisebox{-0.5ex}{ELGs}} & 
\multicolumn{2}{c}{\raisebox{-0.5ex}{ELGs}} &&
\multicolumn{2}{c}{\raisebox{-0.5ex}{ELGs}} & 
\multicolumn{2}{c}{\raisebox{-0.5ex}{ELGs}} & 
\multicolumn{2}{c}{\raisebox{-0.5ex}{ELGs}} &&
\multicolumn{2}{c}{\raisebox{-0.5ex}{ELGs}} & 
\multicolumn{2}{c}{\raisebox{-0.5ex}{ELGs}} & 
\multicolumn{2}{c}{\raisebox{-0.5ex}{ELGs}} \\
 & \multicolumn{2}{c}{(all)} & \multicolumn{2}{c}{(compact)} &
\multicolumn{2}{c}{(diffuse)} &&
\multicolumn{2}{c}{(all)} & \multicolumn{2}{c}{(compact)} &
\multicolumn{2}{c}{(diffuse)} &&
\multicolumn{2}{c}{(all)} & \multicolumn{2}{c}{(compact)} &
\multicolumn{2}{c}{(diffuse)} \\
\hline \hline
R &$-1.1\sigma$ \hspace{-0.95em} & (83) & 
   $\hspace{0.9em}  -2.6\sigma$ \hspace{-0.95em} & (83) \hspace{0.4em} &
     0.9$\sigma$ \hspace{-0.95em} & (83) \hspace{0.4em} &&
    1.0$\sigma$ \hspace{-0.95em} & (53) &
   $\hspace{0.9em} -0.9\sigma$ \hspace{-0.95em} & (53) \hspace{0.4em} &     
    2.1$\sigma$ \hspace{-0.95em} & (53) \hspace{0.4em} &&
   $-1.8\sigma$ \hspace{-0.95em} & (45) & 
   $\hspace{0.9em} -1.3\sigma$ \hspace{-0.95em} & (45) \hspace{0.4em} &
   $-1.1\sigma$ \hspace{-0.95em} & (45) \\
$\Sigma$ & 1.8 & (83) & \hspace{0.9em} 1.9 & (83) \hspace{0.4em} & 
 0.5 & (83) \hspace{0.4em} &&
          $-0.7$ & (53) & \hspace{0.9em} 1.2 & (53) \hspace{0.4em} & 
 $-2.0$ & (53) \hspace{0.4em} &&
           1.7 & (45) & \hspace{0.9em} 1.2 & (45) \hspace{0.4em} & 1.0 & (45) \\  
CT       & 2.7 & (68) & \hspace{0.9em} 4.3 & (68) \hspace{0.4em} & 
 $-0.3$ & (68) \hspace{0.4em} &&
           0.0 & (42) & \hspace{0.9em} 2.5 & (42) \hspace{0.4em} & 
 $-2.6$ & (42) \hspace{0.4em} &&
           3.2 & (37) & \hspace{0.9em} 3.3 & (37) \hspace{0.4em} & 
 1.4 & (37) \\ \hline      
\end{tabular}
\end{table*}

\section {Discussion}
\label{discuss}

\begin{figure*}

\epsfxsize=17cm
\hspace*{0.01cm} \epsffile{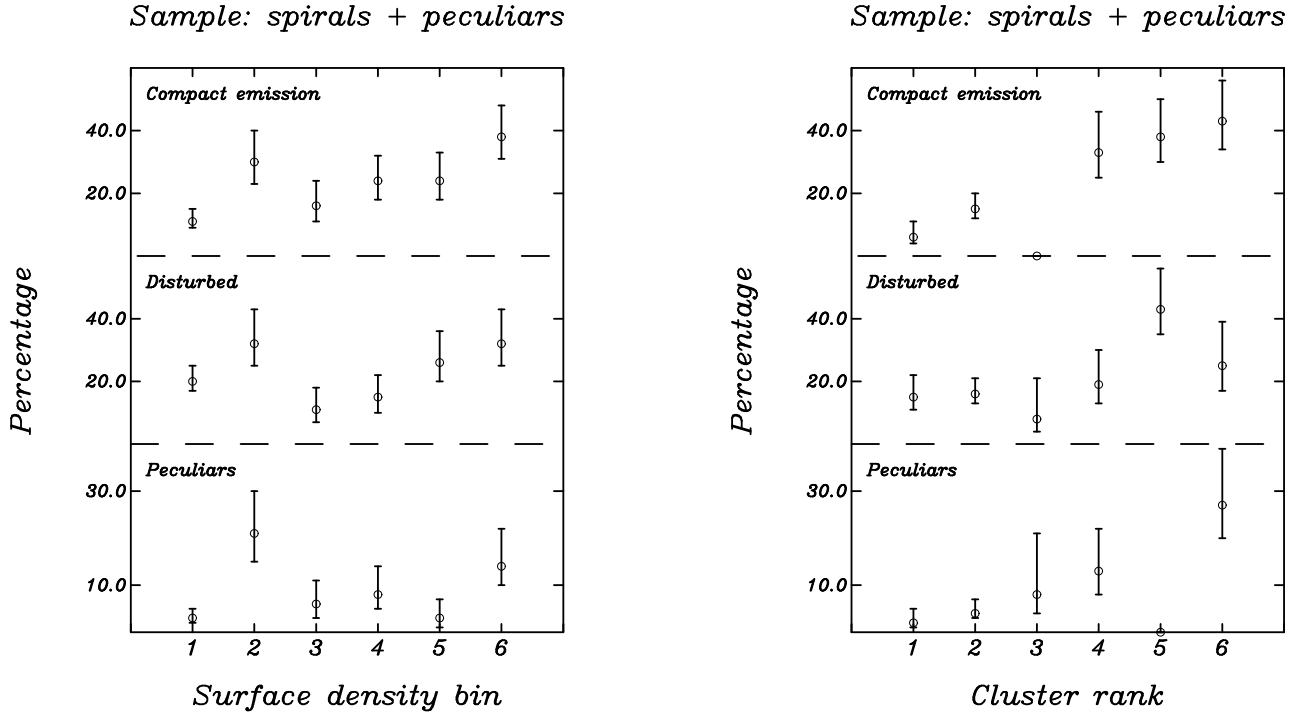}

\caption{\label{scdistrib}
(Left panel) Percentages of galaxies which
have compact emission, are disturbed or are classified as peculiar,
with increasing local galaxy surface density, $\Sigma$.  The surface
density bins, 1--6, correspond to median values of $\Sigma$ equal to
0.0, 0.3, 1.0, 2.3, 4.9 and 14.4 galaxies, $M_{\rm B} \le -19.5$,
Mpc$^{-2}$ respectively. (Right panel) As left panel,
with increasing cluster central galaxy space
density.  Cluster ranks are as follows: 
rank 1, field spirals; rank 2 ($\rho_{1} \sim 2$
Mpc$^{-3}$), Abell 262, 347, 569, 779; rank 3 ($\rho_{1} \sim 3$
Mpc$^{-3}$), Abell 400; rank 4 ($\rho_{1} \sim 4$ Mpc$^{-3}$), Abell
426; rank 5 ($\rho_{1} \sim 7$ Mpc$^{-3}$), Abell 1367; and rank 6
($\rho_{1} \sim 11$ Mpc$^{-3}$), Abell 1656. The galaxy sample for
both panels comprises galaxies classified as spiral or peculiar.
}
\end{figure*}

The analysis of the full cluster sample confirms earlier conclusions
(Papers II and III) that there is an enhancement of tidally-induced
circumnuclear star formation in cluster galaxies (types Sa and later)
compared to similar galaxies in the field.  Whereas previous work
established this simple contrast, the current work shows that the
frequency of circumnuclear starbursts is consistent with a monotonic
increase with increasingly dense cluster environments.  Figure
\ref{scdistrib} shows the increase in the fraction of spirals with
compact emission with cluster rank, from the field (rank 1) to the
richest cluster (Coma, rank 6), as well as with increasing local
galaxy surface density, $\Sigma$.  In particular, we do {\it not}
confirm the result from Hashimoto et al. (1998) who found that poor
clusters have higher levels of starburst emission than either the
field environment or rich clusters.  In fact, the proportion of
spirals with compact emission increases dramatically from the field
($\sim$ 8\%) to the richest cluster (Coma; $\sim$ 43\%).  There are
corresponding increases in the fractions of spirals classed as
peculiar ($\sim$ 2\% in the field; $\sim$ 35\% in Coma), and those
noted as disturbed ($\sim$ 11\% in the field; $\sim$ 39\% and 25\% in
Abell 1367 and Coma respectively).

Is it possible to integrate these findings into a broader picture; one
which addresses cluster evolution from intermediate redshifts to the
present?  Obviously, in a rich cluster such as Coma the residual
spiral fraction is much smaller than the spiral fractions in similar
clusters at intermediate redshift. However, it appears from our study
that {\it the residual spiral population in nearby rich clusters is
similar to the spiral population in clusters at intermediate
redshift}.  The Butcher--Oemler effect would appear to be mainly due to
a {\it decrease} in the spiral population over the last few giga-years,
not primarily a change in the properties of spirals themselves.  The
fraction of spirals in Coma which are peculiar, show signs of
interaction and distortion and which are undergoing tidally-induced
star formation appears similar to the fraction of spirals which show
these effects in rich clusters at $z\sim0.5$.  Yet further evidence
that tides and interactions are important in nearby clusters and not
just in distant clusters has been given by Conselice \& Gallagher
(1999). These authors detect a variety of unusual fine-scale
substructures, including distorted and interacting galaxies, in five
nearby clusters which they consider to be caused by tidal forces.
Trentham \& Mobasher (1998) have discovered a giant
low-surface-brightness arc of length $\sim$ 80 Mpc in the Coma cluster,
and regard fast encounters between nearby galaxies as the likeliest
explanation of its properties.

Lavery \& Henry (1988) first proposed that the Butcher--Oemler effect
could be explained as star formation triggered by galaxy--galaxy
interactions in intermediate redshift clusters. A principal objection
to this hypothesis was that the cluster velocity dispersion is
typically too high ($\sim$ 1000 km s$^{-1}$) for strong tidal
interactions to take place, since these require encounter speeds
comparable to that of the galaxy rotation (Toomre \& Toomre
1972). However, there has been increasing observational evidence for
tidal effects on galaxies both in nearby and intermediate redshift
clusters as well as theoretical work supporting the possibility of
strong tidal fields in clusters. Numerical simulations have shown that
within a few core radii of the centre of a rich cluster such as Coma,
tidal compression of a galaxy by the cluster potential can produce
spiral arms and tidal tails and triggering of enhanced star formation
(e.g. Byrd \& Valtonen 1990; Valluri 1993; Henrikson \& Byrd 1996).
Moore et al. (1996) predict that fast close encounters with the
central massive cluster galaxies will destroy many dwarf galaxies, and
essentially transform spirals into ellipticals or dwarf spheroidals.
All these simulations assume a fixed potential.  However the
potential of a real cluster is expected to vary continually during its
evolution with collapse of the cluster to virialisation, and
subsequent infall of additional material.  Gnedin (1999) has used
self-consistent cluster simulations to demonstrate that this
time-varying potential will cause a sequence of strong tidal shocks on
an individual galaxy, comparable to those from massive galaxies.  The
shocks, which are likely to be produced by surviving groups of
galaxies or large individual galaxies, take place over a wide region
of the cluster, and enhance galaxy--galaxy interactions as well as
amplifying galaxy merger rates.  A galaxy in a cluster similar to that
of Abell richness class 0 or 1 at low redshift, is predicted to have
about 4 encounters closer than 10 kpc per Hubble time, and have a
probability of about 30\% of being in a merger.

These results suggest that profound effects on cluster galaxy
morphology are to be expected from tidal forces during a Hubble time.
In particular, Gnedin demonstrates that a likely consequence of tidal
shocks is to turn a large fraction of normal spirals into S0s. This
occurs by tidal heating of the disk which reduces gravitational
instabilities and suppresses further star formation.  Gas is likely to
be lost by ram-pressure stripping, interpenetrating encounters and,
for low mass galaxies, being blown out by starbursts.  These results thus
suggest that the same tidal forces which we have identified as causing
circumnuclear starbursts in nearby clusters (and are evidently acting
on cluster galaxies at intermediate redshifts), are the primary cause
in transforming the spiral population in distant clusters into the S0
population in present day clusters.

Any mechanism for converting spirals to S0s is required to be more
efficient with increasing galaxy density in order that it could
account qualitatively for the galaxy type--surface density
(T--$\Sigma$) relation found for clusters at $z\sim0.5$ and $z\sim0$
(cf.  Dressler et al. 1997).  As has been seen in section \ref{cfcomp}
above, the frequency of occurrence of tidally-induced starbursts
increases with increasing galaxy surface density, which implies that
tidal forces do indeed act more efficiently on galaxies in higher
density regions. This confirms that these forces are a suitable
mechanism to account, at least qualitatively, for the T--$\Sigma$
relation in clusters.

A further result, obtained in section \ref{cfcomp}, is that the
enhancement of tidally-induced starbursts in cluster spirals is not
wholly accounted for simply by an increase of local galaxy density.
In addition there is a `cluster effect'.  A spiral in a cluster of
higher central galaxy density is more likely to undergo such a
starburst than a spiral in a region of similar local density in a
cluster with a lower central galaxy density.  This implies that, in
regions of comparable local density, the transformation of spirals
into S0s proceeds faster in clusters of higher concentration and/or
higher richness.  This result suggests a simple explanation of the
apparently anomalous absence of a T--$\Sigma$ relation found by
Dressler et al. (1997) for less concentrated, irregular clusters at
intermediate redshifts.

Dressler (1980) had found a significant T--$\Sigma$ relation for
galaxies in both centrally concentrated `regular' clusters and less
concentrated, irregular clusters at $z\sim0$, whereas by contrast
Dressler et al. (1997) found a strong T--$\Sigma$ relation only for regular
clusters at $z\sim0.5$.  Unlike their counterparts at $z\sim0$,
irregular clusters at $z\sim0.5$ have no significant T--$\Sigma$
relation, and ellipticals in these clusters show no concentration to
the densest regions.  This is understandable if there has not
been enough time for a significant transformation of disk galaxy
morphology to take place in irregular clusters at $z\sim0.5$.  By
contrast, such transformation would be expected for regular clusters
at $z\sim0.5$ (for which the timescale for transformation is shorter)
and for irregular clusters at $z\sim0$ (for which a longer time
duration for transformation is available).  Furthermore, the same
galaxy--galaxy and galaxy--group interactions responsible for the
transformation of spirals to S0s may also cause ellipticals to relax
to the densest regions in clusters over similar timescales. Thus our
finding of a `cluster effect' in the enhancement of tidally-induced
starbursts and the consequent inference, for regions of similar local
density, of an accelerated transformation of spirals to S0s in
clusters of higher central density together provide a natural
explanation for the apparently anomalous absence of a T--$\Sigma$ relation for
galaxies in irregular clusters at $z\sim0.5$.

Finally, one may ask what mechanism may accelerate the rate of galaxy
encounters (and consequent starburst activity), in clusters with
greater central galaxy density?  The work by Gnedin (1999) has shown
that a time varying cluster potential will enhance such encounters.
Such a varying potential will arise both from cluster infall, and from
subcluster mergers.  Recent X-ray studies of a number of clusters have
shown asymmetric X-ray morphologies and temperature structures which
are consistent with those seen in simulations of subcluster mergers
(e.g. Henriksen \& Markevitch 1996; Donnelly et al. 1998; Honda et
al. 1996; Henriksen, Wang \& Ulmer 1999), implying that these clusters
are recent postmerger systems.  Furthermore, from a study of 10
distant clusters, Wang \& Ulmer (1999) have shown that cluster global
X-ray ellipticities correlate with their blue galaxy fractions.  The
strongly elongated clusters show substantial amounts of substructure,
indicating that they are dynamically young systems, and leading Wang
\& Ulmer to suggest that the blue cluster galaxies originate in the
process of cluster formation.

The above results thus suggest that subcluster mergers may be a
mechanism to drive an accelerated rate of galaxy encounters and
tidally-induced starbursts (and consequent morphological evolution of
disk galaxies) in more centrally concentrated clusters.  One may
suppose that such clusters have formed either as a result of
subcluster mergers, or in higher density regions where the probability
of such subcluster accretion is greater.  The consequent accelerated
rate of galaxy encounters and morphological evolution would account
for a significant T--$\Sigma$ relation for these clusters at $z \sim
0.5$ as compared to the absence of such a relation for the
(presumably) relatively isolated irregular clusters at the same
redshift.

According to this picture, a significant enhancement of starburst
activity above that normally expected for galaxies in a region of a
given density, is expected in clusters which are still undergoing the
effects of subcluster merger.  By contrast no such enhancement would
be evident for clusters which are more relaxed.  Such a scenario is
entirely consistent with the results of our survey.  The two most
centrally concentrated clusters in the survey, Abell 1367 and Coma
both show evidence of being recent postmerger systems (Donnelly et
al. 1998; Honda et al. 1996). And, in accord with the expectation for
such systems, spiral galaxies in these clusters have been found to
have an enhanced starburst activity as compared to spirals in regions
of similar density in less concentrated clusters.

\section {Conclusions}
\label{conclude}

From a survey of H$\alpha$ emission in galaxies of types Sa and later
in 8 low-redshift Abell clusters, we have shown that circumnuclear
starbursts, most probably triggered by tidal interactions
(galaxy--galaxy, galaxy--group and galaxy--cluster), are more
prevalent in spirals in denser regions and in clusters with a greater
central galaxy density. In contrast to previous work, we find a
monotonic increase in the fraction of spirals undergoing these
starbursts from the field to higher density regions, and from clusters
with low central galaxy density to clusters with high central
density. There is a similar increase in the fraction of spirals
classified as disturbed between the field and higher density environments,
and between clusters of low and high central density.
In the richest cluster studied (Coma), the fraction of spirals undergoing
tidal distortion and/or tidally-induced star formation appears 
comparable to the fraction of spirals showing these effects in rich
clusters at $z\sim0.5$.

From these results it is suggested that tidal interactions are the
primary mechanism for an on-going transformation of spirals to S0s in
clusters, a scenario fully in accord with the most recent models of
clusters with a non-static potential undergoing collapse and
infall. This mechanism can qualitatively account for the type--local
surface density (T--$\Sigma$) relation found in clusters on account of
the higher efficiency of the mechanism in higher density
regions. Furthermore the prevalance of tidally-induced starbursts in
spirals is found to depend not solely on local galaxy density, but
also on cluster type.  This implies that, for regions of comparable
local density, transformation of spirals to S0s will take place faster
in clusters with higher central density.  This can account for the
apparently anomalous lack of a T--$\Sigma$ relation for irregular
clusters at intermediate redshift.  For these clusters there has not
been time for significant morphological transformation of disk
galaxies to have taken place in contrast to regular clusters at
intermediate redshift (for which the timescale for transformation is
shorter) and for low-redshift irregular clusters (for which a longer
time interval is available during which transformations may take
place).  Moreover it is suggested that subcluster merging is a cause
of the enhanced starburst activity (and consequent accelerated
morphological evolution of disk galaxies) seen in the denser clusters,
as compared to regions of similar density in less dense clusters.  The
two richest clusters in our survey show evidence of being recent
postmerger systems, whose galaxies have such enhanced starburst
activity, consistent with this picture.
  
Finally, the fraction of late-type galaxies which are classified as
peculiar (i.e.  not in a recognisable stage of the Hubble sequence)
also increases from the field to higher density environments, and from
clusters of low to higher central density, in parallel with the
increasing prevalance of tidally-induced starbursts in spirals. A very
high fraction ($\sim$ 70\%) of these galaxies have emission similar to
the starburst emission of spirals.  It is suggested that these
galaxies are predominantly on-going mergers, which are expected as the
end product of some of the tidal interactions, and which are expected
to be more common in regions of higher density and in clusters of
higher central density due to the greater prevalence of tidal
interactions in these locations.

\section*{Acknowledgements}

CM and MW thank the Institute of Astronomy, Cambridge, and CM thanks
the Department of Astronomy, University of Virginia for hospitality in
the course of this project.  
We would like to thank A. Biviano for helpful comments, and for kindly 
providing a Fortran code to estimate partial rank correlation coefficients
using a bootstrap resampling technique.
Observations were made with the Burrell
Schmidt telescope of the Warner and Swasey Observatory, Case Western
Reserve University. This research has made use of the NASA/IPAC
Extragalactic Database (NED) which is operated by the Jet Propulsion
Laboratory, California Institute of Technology, under contract with
the National Aeronautics and Space Administration.


\begin{thebibliography}{}

\bibitem[\protect\citename{Abell  }1958]{abe} 

Abell G. O., 1958, ApJS, 3, 211


\bibitem[\protect\citename{Abell et al.  }1989]{aco}

Abell G.O., Corwin H.G., Olowin R.P., 1989, ApJS, 70, 1


\bibitem[\protect\citename{Balogh et al.  }1998]{ba98}

Balogh M.L., Schade D., Morris S.L., Yee H.K.C., Carlberg R.G.,
Ellingson E., 1998, ApJ, 504, L75


\bibitem[\protect\citename{Balzano  }1983]{bal83}

Balzano V.A., 1983, ApJ, 268, 602


\bibitem[\protect\citename{Baugh et al.  }1996]{bau96}

Baugh C.M., Cole S., Frenk C.S., 1996, MNRAS, 283, 1361


\bibitem[\protect\citename{Bennett \& Moss  }1998]{bm98}

Bennett S.M., Moss C., 1998, A\&AS, 132, 55


\bibitem[\protect\citename{Biviano et al.   }1997]{biv97}

Biviano A., Katgert P., Mazure A., Moles M., den Hartog R.,
Perea J., Focardi P., 1997, A\&A, 321, 84


\bibitem[\protect\citename{Bushouse  }1987]{bus87}

Bushouse H.A., 1987, ApJ, 320, 49


\bibitem[\protect\citename{Butcher \& Oemler  }1978]{bo78}

Butcher H., Oemler A., 1978, ApJ, 219, 18


\bibitem[\protect\citename{Byrd \& Valtonen  }1990]{byrv90}

Byrd G., Valtonen M., 1990, ApJ, 350, 89


\bibitem[\protect\citename{Conselice \& Gallagher  }1999]{cong99}

Conselice C.J., Gallagher J.S., 1999, AJ, 117, 75


\bibitem[\protect\citename{Cowie \& Songaila  }1977]{cows77}

Cowie L.L., Songaila A., 1977, Nat, 266, 501


\bibitem[\protect\citename{de Jong  }1995]{jong95}

de Jong R.S., 1995, PhD thesis. Univ. Groningen


\bibitem[\protect\citename{de Vaucouleurs  }1959]{gv59}

de Vaucouleurs G., 1959, in Fl\"{u}gge, S., ed., Handbuch der Physik,
Vol. LIII. Springer-Verlag, Berlin, p. 275


\bibitem[\protect\citename{de Vaucouleurs  }1974]{gv74}

de Vaucouleurs G., 1974, in Shakeshaft, J.R., ed.,
The Formation and Dynamics of Galaxies,
IAU Symposium No. 58. Reidel, p.1


\bibitem[\protect\citename{de Vaucouleurs et al. }1976]{rc2}

de Vaucouleurs G., de Vaucouleurs A., Corwin H.G., 1976, Second Reference
Catalogue of Bright Galaxies. University of Texas Press, Austin


\bibitem[\protect\citename{de Vaucouleurs et al.  }1991]{rc3}

de Vaucouleurs G., de Vaucouleurs A., Corwin H.G., Buta R.J.,
Paturel G., Fouqu\'{e} P., 1991, Third Reference Catalogue of Bright
Galaxies. Springer-Verlag, New York


\bibitem[\protect\citename{Donnelly et al.  }1998]{don98}

Donnelly R.H., Markevitch M., Forman W., Jones C., David L.P.,
Churazov E., Gilfanov M., 1998, ApJ, 500, 138 


\bibitem[\protect\citename{Donas et al.  }1990]{don90}

Donas J, Milliard B., Laget M., Buat V., 1990, A\&A, 235,60



\bibitem[\protect\citename{Dressler  }1980]{dr80}

Dressler A., 1980, ApJ, 236, 351


\bibitem[\protect\citename{Dressler et al.  }1994]{dr94}

Dressler A., Oemler A., Butcher H.R., Gunn J.E., 1994,
ApJ, 430, 107


\bibitem[\protect\citename{Dressler et al.  }1997]{dr97}

Dressler A. et al., 1997, ApJ, 490, 577

\bibitem[\protect\citename{Gavazzi \& Contursi  }1994]{gavc94}

Gavazzi G., Contursi A., 1994, AJ, 108, 24


\bibitem[\protect\citename{Gavazzi et al.  }1996]{gpb}

Gavazzi G., Pierini D., Boselli A., 1996, A\&A, 312, 397


\bibitem[\protect\citename{Gavazzi et al.  }1998]{gcc}

Gavazzi G., Catinella B., Carrasco L., Boselli A., Contursi A., 1998,
ApJ, 115, 1745


\bibitem[\protect\citename{Gnedin  }1999]{gn99}

Gnedin O.Y., 1999, PhD thesis, Princeton Univ.


\bibitem[\protect\citename{Gunn \& Gott  }1972]{gug72}

Gunn J.E., Gott J.R., 1972, ApJ, 176, 1


\bibitem[\protect\citename{Hameed \& Devereux  }1999]{hd99}

Hameed S., Devereux N., 1999, AJ, 118, 730


\bibitem[\protect\citename{Hashimoto et al.  }1998]{hash98}

Hashimoto Y., Oemler A., Lin H., Tucker D.L., 1998, ApJ, 499, 589


\bibitem[\protect\citename{Heckman  }1980]{heck80}

Heckman T.M., 1980, A\&A, 87, 152


\bibitem[\protect\citename{Henriksen \& Byrd  }1996]{hb96}

Henriksen M.J., Byrd G., 1996, ApJ, 459, 82


\bibitem[\protect\citename{Henriksen \& Markevitch  }1996]{hm96}

Henriksen M.J., Markevitch M.L., 1996, ApJ, 466, L79


\bibitem[\protect\citename{Henriksen et al.  }1999]{hwu99}

Henriksen M.J., Wang Q.D., Ulmer M., 1999, MNRAS, 307, 67


\bibitem[\protect\citename{Hernquist  }1989]{hern89}

Hernquist L., 1989, Nat, 340, 687


\bibitem[\protect\citename{Ho et al.  }1997a]{ho97a}

Ho L.C., Filippenko A.V., Sargent W.L.W., 1997a, ApJ, 487, 568


\bibitem[\protect\citename{Ho et al.  }1997b]{ho97b}

Ho L.C., Filippenko A.V., Sargent W.L.W., 1997b, ApJ, 487, 591


\bibitem[\protect\citename{Honda et al.  }1996]{hon96}

Honda H. et al., 1996, ApJ, 473, L71


\bibitem[\protect\citename{Hubble \& Humason }1931]{hh31}

Hubble E., Humason M.L., 1931, ApJ, 74, 43


\bibitem[\protect\citename{Huchra et al.  }1995]{cfa}

Huchra J.P., Geller M.J., Clemens C.M., Tokarz S.P., Michel A., 1995,
The CfA Redshift Catalogue, Version June 1995.


\bibitem[\protect\citename{Keel et al.  }1985]{keel85}

Keel W.C., Kennicutt R.C., Hummel E., van der Hulst J.M., 1985,
AJ, 90, 708


\bibitem[\protect\citename{Kennicutt }1992]{k92}

Kennicutt R.C., 1992, ApJ, 388, 310


\bibitem[\protect\citename{Kennicutt }1998]{k98}

Kennicutt R.C., 1998, ARA\&A, 36, 189


\bibitem[\protect\citename{Kennicutt \& Kent  }1983]{kk83}

Kennicutt R.C., Kent S.M., 1983, AJ, 88, 1094


\bibitem[\protect\citename{Kennicutt et al.  }1987]{kenn87}

Kennicutt R.C., Roettiger K.A., Keel W.C., van der Hulst J.M.,
Hummel E., 1987, AJ, 93, 1011


\bibitem[\protect\citename{Koopmann \& Kenney  }1998]{kok98}

Koopmann R.A., Kenney J.D.P., 1998, ApJ, 497, L75


\bibitem[\protect\citename{Lavery \& Henry  }1988]{lh88}

Lavery R.J., Henry J.P., 1988, ApJ, 330, 596


\bibitem[\protect\citename{Mihos \& Hernquist  }1996]{mihh96}

Mihos J.C., Hernquist L., 1996, ApJ, 464, 641


\bibitem[\protect\citename{Miller  }1988]{mill88}

Miller R.H., 1988, Comm.Ap., 13, 1


\bibitem[\protect\citename{Moore et al.  }1996]{moo96}

Moore B., Katz N., Lake G., Dressler A., Oemler A., 1996,
Nat, 379, 613


\bibitem[\protect\citename{Moss \& Whittle  }1993]{mw93}

Moss C., Whittle M., 1993, ApJ, 407, L17 (Paper II)


\bibitem[\protect\citename{Moss et al.  }1988]{mwi}

Moss C., Whittle M., Irwin M.J., 1988, MNRAS, 232, 381 (Paper I)


\bibitem[\protect\citename{Moss et al.  }1998]{mwp}

Moss C., Whittle M., Pesce J.E., 1998, MNRAS, 300, 205 (Paper III) 


\bibitem[\protect\citename{Nilson  }1973]{nil} 

Nilson P., 1973, Uppsala General Catalogue of Galaxies, 
Uppsala astr. Obs. Ann., 6


\bibitem[\protect\citename{Noguchi  }1988]{no88} 

Noguchi M., 1988, A\&A, 203, 259


\bibitem[\protect\citename{Noguchi \& Ishibashi  }1986]{nois86} 

Noguchi M., Ishibashi S., 1986, MNRAS, 219, 305


\bibitem[\protect\citename{Oemler    }1974]{oem74}

Oemler A., 1974, ApJ, 194, 1


\bibitem[\protect\citename{Oemler et al.   }1997]{oem97}

Oemler A., Dressler A., Butcher H.R.,  1997, ApJ, 474, 561


\bibitem[\protect\citename{Paturel et al.  }1994]{pbg}

Paturel G., Bottinelli L., Gouguenheim L., 1994, A\&A, 286, 768


\bibitem[\protect\citename{Romanishin  }1990]{r90}

Romanishin W., 1990, AJ, 100, 373

\bibitem[\protect\citename{Sandage \& Tammann  }1987]{st87} 

Sandage A., Tammann G.A., 1987, A Revised Shapley-Ames Catalog of Bright
Galaxies, Carnegie Institution of Washington Publ. 635


\bibitem[\protect\citename{Sanders et al.   }1988]{san88} 

Sanders D.B., Soifer B.T., Elias J.H., Madore B.F., Matthews K.,
Neugebauer G., Scoville N.Z., 1988, ApJ, 325, 74


\bibitem[\protect\citename{Simien \& de Vaucouleurs  }1986]{sivau86}

Simien F., de Vaucouleurs G., 1986, ApJ, 302, 564


\bibitem[\protect\citename{Smail et al.  }1997]{sm97}

Smail I., Dressler A., Couch W.J., Ellis R.S., Oemler A.,
Butcher H., Sharples R.M., 1997, ApJS, 110, 213


\bibitem[\protect\citename{Spitzer \& Baade  }1951]{spb51}

Spitzer L., Baade W., 1951, ApJ, 113, 413


\bibitem[\protect\citename{Strauss et al.  }1992]{str}

Strauss M.A., Huchra J.P., Davis M., Yahil A., Fisher K.B., 
Tonry J., 1992, ApJS, 83, 29 


\bibitem[\protect\citename{Struble \& Rood  }1991]{sr} 

Struble M. F., Rood H. J., 1991, ApJS, 77, 363


\bibitem[\protect\citename{Toomre \& Toomre  }1972]{tt72} 

Toomre A., Toomre J., 1972, ApJ, 178, 623


\bibitem[\protect\citename{Trentham \& Mobasher  }1998]{trmo98} 

Trentham N., Mobasher B., 1998, MNRAS, 293, 53


\bibitem[\protect\citename{Valluri  }1993]{vall93} 

Valluri M., 1993, ApJ, 408, 57


\bibitem[\protect\citename{Valluri \& Jog  }1990]{vjog90} 

Valluri M., Jog C.J., 1990, ApJ, 357, 367


\bibitem[\protect\citename{van den Bergh  }1976]{vdb76} 

van den Bergh S., 1976, ApJ, 206, 883


\bibitem[\protect\citename{van den Bergh  }1997]{vdb97} 

van den Bergh S., 1997, AJ, 113, 2054


\bibitem[\protect\citename{Wallenquist  }1960]{wa60} 

Wallenquist A., 1960, Medd. Uppsala Astr. Obs., No. 127


\bibitem[\protect\citename{Wang \& Ulmer  }1997]{wu97} 

Wang Q.D., Ulmer M.P., 1997, MNRAS, 292, 920


\bibitem[\protect\citename{Wright et al.  }1988]{wri88} 

Wright G.S., Joseph R.D., Robertson N.A., James P.A., Meikle W.P.S.,
1988, MNRAS, 233, 1


\bibitem[\protect\citename{Zwicky et al.  }1960--1968]{zwi}

Zwicky F., Herzog E., Wild P., Karpowicz M., Kowal C., 1960--1968,
Catalogue of Galaxies and Clusters of Galaxies, Vols. 1--6.
California Inst. Tech., Pasadena


\end{thebibliography}
\end{document}